\documentclass[submission, Phys]{SciPost}

\usepackage[utf8]{inputenc} 
\usepackage[T1]{fontenc} 	
\usepackage[english]{babel} 

\usepackage[bitstream-charter]{mathdesign}
\usepackage{geometry} 		
\usepackage{amsmath} 		
\usepackage{mathtools} 		
\usepackage{float} 			
\usepackage{graphicx} 		
\usepackage{tabularx} 		
\usepackage{booktabs} 		
\usepackage{color, xcolor} 	
\usepackage{pdfpages} 		
\usepackage{extarrows} 		
\usepackage{multirow} 		
\usepackage{multicol} 		
\usepackage{subcaption} 	
\usepackage{enumitem} 		
\usepackage{xspace} 		
\usepackage{stackrel} 		
\usepackage{tikz} 			
\usepackage{braket} 		
\usepackage{bm} 			
\usepackage{tensor} 		
\usepackage{slashed} 		
\usepackage{siunitx} 		
\usepackage{lastpage} 		
\usepackage{cite} 			
\usepackage[normalem]{ulem} 
\usepackage{fontawesome} 	
\usepackage{tocloft} 		
\usepackage{titlesec} 		
\usepackage{doi} 			
\usepackage{hyperref} 		
\usepackage[most]{tcolorbox} 					
\usepackage[nameinlink, capitalize]{cleveref} 	
\usepackage[nottoc, notlot, notlof]{tocbibind} 	
\usepackage[ruled, vlined]{algorithm2e} 		
\usepackage{makecell}
\usepackage[makeroom]{cancel}
\usepackage{feynmf}

\usepackage{multicol}

\makeatletter
\def\BState{\State\hskip-\ALG@thistlm}
\makeatother

\makeatletter
\@ifundefined{pdfoutput}{}{\DeclareGraphicsRule{*}{mps}{*}{}}
\makeatother

\makeatletter
\DeclareRobustCommand*{\bfseries}{%
   \not@math@alphabet\bfseries\mathbf
   \fontseries\bfdefault\selectfont
   \boldmath
}
\makeatother

\hypersetup{
	pdftitle={CUDACPP},
	pdfauthor={Wettersten et al.},
	colorlinks=true, 			
	linkcolor={red!50!black}, 	
	citecolor={blue!50!black}, 	
	urlcolor={blue!80!black} 	
} 

\DeclareSymbolFont{usualmathcal}{OMS}{cmsy}{m}{n}
\DeclareSymbolFontAlphabet{\mathcal}{usualmathcal}

\SetArgSty{textnormal}
\SetKwComment{Comment}{{\small\#}~}{}
\SetCommentSty{mycommfont}

\setitemize{itemsep=0pt, parsep=0pt} 				
\setenumerate{itemsep=0pt, parsep=0pt} 				
\setitemize{itemsep=2pt,topsep=2pt,parsep=0pt,partopsep=0pt,leftmargin=*}
\setenumerate{itemsep=0pt,topsep=2pt,parsep=0pt,partopsep=0pt,labelindent=3pt,leftmargin=*}
\usepackage{amsmath}
 
\usepackage{amsthm} 		
\theoremstyle{definition}
 
\definecolor{red_cb}{HTML}{e41a1c}
\definecolor{blue_cb}{HTML}{377eb8}
\definecolor{green_cb}{HTML}{4daf4a}
\definecolor{purple_cb}{HTML}{984ea3}
\definecolor{orange_cb}{HTML}{ff7f00}

\definecolor{EmeraldGreen}{HTML}{1ea78d}
\definecolor{EnglishRed}{HTML}{b02427}
\hypersetup{colorlinks=true,urlcolor=EmeraldGreen,citecolor=EmeraldGreen,linkcolor=EnglishRed}

\newcommand{\cudacpp}{\texttt{CUDACPP}}
\newcommand{\vecsize}{\texttt{VECSIZE}}

\newcommand\one{\leavevmode\hbox{\small1\normalsize\kern-.33em1}}

\newcommand{\madgraph}{\textsc{MadGraph5\_aMC@NLO}\xspace}
\newcommand{\mg}{\textsc{MG5aMC}\xspace}

\newcommand{\arXiv}[2][]{%
	\ifthenelse{\equal{#1}{}}%
	{\href{http://arxiv.org/abs/#2}{arXiv:#2}}%
	{\href{http://arxiv.org/abs/#2}{arXiv:#2~[#1]}}}

\def\slashchar#1{\setbox0=\hbox{$#1$}           
   \dimen0=\wd0                                 
   \setbox1=\hbox{/} \dimen1=\wd1               
   \ifdim\dimen0>\dimen1                        
      \rlap{\hbox to \dimen0{\hfil/\hfil}}      
      #1                                        
   \else                                        
      \rlap{\hbox to \dimen1{\hfil$#1$\hfil}}   
      /                                         
   \fi}

\newcommand{\tikznode}[2]{%
\ifmmode%
\tikz[remember picture,baseline=(#1.base),inner sep=0pt] \node (#1) {$#2$};%
\else
\tikz[remember picture,baseline=(#1.base),inner sep=0pt] \node (#1) {#2};%
\fi}

\def\mathswitchr#1{\relax\ifmmode{\mathrm{#1}}\else$\mathrm{#1}$\xspace\fi}
\def\mathswitch#1{\relax\ifmmode#1\else$#1$\xspace\fi}

\usepackage{cleveref}
\usepackage{soul}
\usepackage{verbatim}
\usepackage{listings}
\usepackage[normalem]{ulem}

\lstdefinestyle{numbers} {numbers=left, stepnumber=1, numberstyle=\tiny, numbersep=10pt}

\lstdefinelanguage{madgraph}
{
    keywords=[1]{
    CALL, DO, END
    },
    keywordstyle=[1]\color[HTML]{228B22},
    keywords=[2]{
        VXXXXX, OXXXXX, IXXXXX, VVV1P0_1, FFV1_0
    },
    keywordstyle=[2]\color[HTML]{1027ad},
    basicstyle=\ttfamily\linespread{4},
    breaklines=false,
    columns=flexible,
    commentstyle=\color[rgb]{0.127,0.427,0.514}\ttfamily\itshape,
    escapechar=@,
    extendedchars=true,
    inputencoding=latin1,
    numbers=left,
    numberstyle=\tiny,
    comment=[l]{!},
    numbers=left,
    numberstyle=\tiny,
}

\lstdefinelanguage{cudacpp}
{
    keywords=[1]{
    M_ACCESS, W_ACCESS, CD_ACCESS, A_ACCESS 
    },
    keywordstyle=[1]\color[HTML]{228B22},
    keywords=[2]{
        vxxxxx, oxxxxx, ixxxxx, VVV1P0_1, FFV1_0
    },
    keywordstyle=[2]\color[HTML]{1027ad},
    basicstyle=\ttfamily\linespread{4},
    breaklines=false,
    columns=flexible,
    commentstyle=\color[rgb]{0.127,0.427,0.514}\ttfamily\itshape,
    escapechar=@,
    extendedchars=true,
    inputencoding=latin1,
    numbers=left,
    numberstyle=\tiny,
    comment=[l]{//},
    numbers=left,
    numberstyle=\tiny,
}

\lstdefinelanguage{terminal}
{
    alsoletter={._},
    keywords=[1]{
    cd, ln, git, generate, output, make, check_cpp.exe, launch, install
    },
    keywordstyle=[1]\color[HTML]{228B22},
    keywords=[2]{
        clone, s, BACKEND, FPTYPE, USEOPENMP, standalone_cudacpp, madevent_simd, madevent_gpu
    },
    keywordstyle=[2]\color[HTML]{1027ad},
    basicstyle=\ttfamily\linespread{4},
    breaklines=false,
    columns=flexible,
    commentstyle=\color[rgb]{0.127,0.427,0.514}\ttfamily\itshape,
    escapechar=@,
    extendedchars=true,
    inputencoding=latin1,
    comment=[l]{\#},
    literate={.git}{.git}4,
}

\graphicspath{{./figs/}}

\begin{document}

\newcommand{\om}[1]{{\color{red}{#1}}}

\begin{center}{\Large \textbf{Data-parallel leading-order event generation in \madgraph{}}}
\end{center}

\begin{center}
Stephan Hageb{\"o}ck\textsuperscript{1},
Daniele Massaro\textsuperscript{1},
Olivier Mattelaer\textsuperscript{2},
Stefan Roiser\textsuperscript{1},
Andrea Valassi\textsuperscript{1},
and Zenny Wettersten \textsuperscript{1,3}
\end{center}

\begin{center}
{\bf 1} CERN, Geneva, Switzerland \\
{\bf 2} CP3, Universit\'e catholique de Louvain, Louvain-la-Neuve, Belgium \\
{\bf 3} HEPHY, {\"O}AW, Vienna, Austria
\end{center}

\begin{center}
\today
\end{center}


\section*{Abstract}
The \cudacpp{} plugin for \madgraph{} aims to accelerate leading order tree-level event generation by providing the \textsc{MadEvent} event generator with data-parallel helicity amplitudes. These amplitudes are written in templated C++ and CUDA, allowing them to be compiled for CPUs supporting SSE4, AVX2, and AVX-512 instruction sets as well as CUDA- and HIP-enabled GPUs. Using SIMD instruction sets, \cudacpp{}-generated amplitude routines are shown to speed up linearly with SIMD register size, and GPU offloading is shown to provide acceleration beyond that of SIMD instructions. Additionally, the resulting speed-up in event generation perfectly aligns with predictions from measured runtime fractions spent in amplitude routines, and proper GPU utilisation can speed up high-multiplicity QCD processes by an order of magnitude when compared to optimal CPU usage in server-grade CPUs.

\vspace{10pt}
\noindent\rule{\textwidth}{1pt}
\tableofcontents\thispagestyle{fancy}
\noindent\rule{\textwidth}{1pt}
\vspace{10pt}

\clearpage
\section{Introduction}
\label{sec:intro}

With the approaching High-Luminosity LHC upgrade (HL-LHC) \cite{Schmidt_2016}, there are several technical challenges that need to be overcome in order to make the upgrade a success. One such challenge is the increasing computing resource requirements, which the projected resources of the Worldwide LHC Computing Grid (WLCG) will not be able to provide without major improvement of the software stack \cite{HEPSoftwareFoundation:2017ggl}. While event reconstruction (using Geant4 \cite{ALLISON2016186}) is the main compute bottleneck, Monte-Carlo (MC) event generation is expected to have a significant compute cost in the HL-LHC era, on the order of $10-20\%$ of total CPU usage for the ATLAS and CMS experiments \cite{CERN-LHCC-2022-005,Software:2815292}, and substantial speed-up is therefore required here as well \cite{HSFPhysicsEventGeneratorWG:2020gxw}.
The goal of this paper is therefore --- in the context of a specific tool --- to accelerate MC event generation from the perspective of hardware utilisation by rewriting the program to properly utilise both available CPU resources and enable GPU offloading. 

MC event generation is the stochastic sampling of a quantum field theory (QFT) into individual potential interactions --- events --- that are used for statistical predictions in a given QFT model. Many tools \cite{Buckley:2011ms,Sjostrand:2016bif,Campbell:2022qmc} have been created over the years to tackle such stochastic sampling. In order to speed up this type of simulation, different software methods and hardware architectures need to be combined: for example, many collaborations have in recent years put in effort to improve event generation by e.g.\ reducing the fraction of negatively weighted events in next-to-leading order (NLO) samples \cite{Frederix:2020trv,Andersen:2021mvw,Frederix:2023hom}, optimising event generation algorithms \cite{Mattelaer:2021xdr,Lifson:2022mxf}, studying better representations of the underlying theoretical calculations \cite{Hagiwara:2020tbx,Chen:2022gxv,Chen:2022xlg,Hagiwara:2024xdh,Lifson:2022ijv,Boman:2023afu}, and particularly accelerating event generators with hardware acceleration \cite{Kanzaki:2010ym,Hagiwara:2010oca,Hagiwara:2013oka,Bothmann:2021nch,Carrazza:2021gpx,Bothmann:2023gew,Cruz-Martinez:2025kwa,Valassi:2021ljk,Valassi:2022dkc,Valassi:2023yud,Hageboeck:2023blb,Valassi:2025xfn}.  

\madgraph{} (\mg{}) \cite{Alwall:2014hca} is a general-purpose high-energy physics (HEP) toolkit for Standard Model (SM) and beyond-the-SM (BSM) phenomenology, for example facilitating  the generation of unweighted parton-level events at leading order (LO) and NLO+parton shower (PS) accuracy, and is used extensively by LHC experiments as part of their simulation chains. Due to the non-branching nature of (leading-order) scattering amplitudes and the significant amount of phase-space points these scattering amplitudes need to be evaluated at, parton-level event generation makes an optimal candidate for hardware acceleration using data-parallel hardware architectures with single instruction across multiple data (SIMD) instruction sets (available in most modern CPUs), and single instruction across multiple threads (SIMT) architectures (such as general-purpose GPUs). Based on this motivation there have been several attempts at running \mg{} event generation on GPUs by porting scattering amplitude code to CUDA \cite{Kanzaki:2010ym,Hagiwara:2010oca,Hagiwara:2013oka} and using automated deployment libraries \cite{Carrazza:2021gpx}.

As the original work on GPU offloading \cite{Kanzaki:2010ym,Hagiwara:2010oca,Hagiwara:2013oka} never reached production, this effort was restarted in 2020 in the face of the upcoming computing requirements mentioned above; this new project has resulted in the \cudacpp{} plugin presented in this paper. \cudacpp{} offers an alternative output mode in \mg{} to generate explicitly vectorised C++ code and CUDA code (in lieu of the ALOHA-generated \cite{deAquino:2011ub} Fortran code exported by default) for LO processes. Furthermore, this code has been templated so as to allow compilation for scalar, SSE4, AVX2, and AVX-512 instruction sets, and both Nvidia and AMD GPUs. Only the scattering amplitudes are considered here; however, as Feynman diagram methods have factorial complexity scaling with the number of external partons, scattering amplitudes quickly become the dominant runtime bottleneck as processes grow more complex. The surrounding event generation infrastructure is identical to the upstream \mg{} project, although some changes were required and have been integrated into the upstream \mg{} codebase.

This paper is structured as follows: \cref{sec:hardware,sec:datapara_evgen} provide brief introductions to hardware and software specifications, respectively, with the former including a brief overview of SIMD and SIMT data parallelism in the context of high-performance computing (HPC) and the latter a simple introduction to the \textsc{MadEvent} structure and the specific \cudacpp{} implementations. \Cref{sec:standalone} details results for \cudacpp{} amplitudes; primarily the increase in throughput (defined as the number of scattering amplitudes evaluated per second), but also a study on the needs for higher precision in helicity amplitudes and tests of \cudacpp{}'s ``physically motivated'' FP64-/FP32-mixed precision mode where kinematic structures are evaluated in FP64 while colour algebra is performed in FP32. The real-world \cudacpp{} event generation speed-up is shown in \cref{sec:evgen}, where runtimes to generate varying amounts of unweighted events in pre-compiled gridpacks using both SIMD CPUs\footnote{Here, SIMD CPUs refer to general-purpose CPUs with hardware-level SIMD instructions. These are often called vector or vectorised CPUs, but to avoid confusion with the generic term ``vector processor'' the term SIMD CPU is used throughout this text.} and SIMT GPUs are measured. Finally, \cref{sec:conclusions} summarises the results and provides discussion in the context of surrounding concurrent research, alongside an outlook into upcoming developments both within the context of \cudacpp{} itself as well as related research done by other groups and external collaborators. Additionally, a quick-start manual to install and use \cudacpp{} is given in \cref{appendix:manual}.

\section{Hardware specifications: SIMD CPUs and SIMT GPUs}
\label{sec:hardware}

Although the general considerations for data-parallel software coincide across hardware architectures, the specifics are important to ensure optimal performance on a given system. The architectures relevant here are SIMD CPUs and SIMT GPUs, both of which perform the same instruction across multiple data streams, but do so at incredibly different scales and with considerably different restrictions. These hardware implementations, their advantages, and their restrictions are presented briefly in \cref{sec:simd,sec:simt} respectively.

\subsection{Single instruction, multiple data}
\label{sec:simd}

\begin{figure}[t]
    \centering
    \includegraphics[height=3.2cm]{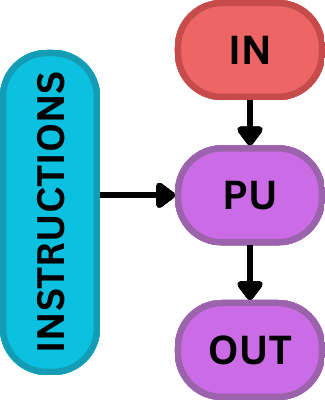}
    \hspace{1.4cm}
    \includegraphics[height=3.2cm]{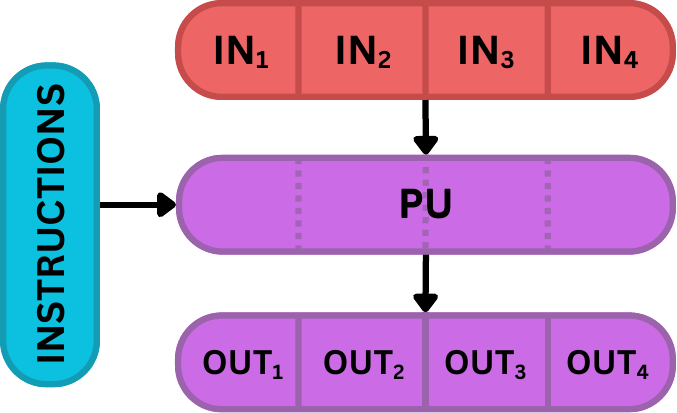}
    \caption{Simple illustration of the difference between SISD (left) and SIMD (right) processing. In both cases, instructions and input data are loaded into a processing unit (PU) which outputs the modified data. In a SIMD PU, though, the data register is sufficiently large that several data can be loaded next to each other, and the corresponding SIMD instruction is implemented as the corresponding scalar instruction acting against each ``part'' of the data register independently.}
    \label{fig:sisd_vs_simd}
\end{figure}

SIMD processing implements hardware-level data parallelism as a minimal extension to classical scalar (single instruction, single datum; SISD) CPU architectures. Scalar CPUs load a single instruction and a single datum into a processing unit (PU), which then outputs the resulting modified datum. By extending the data register size of the PU to a multiple of the size of standard data types, a SIMD instruction can be implemented as the application of the corresponding scalar instruction evenly across the data. This is graphically represented in \cref{fig:sisd_vs_simd}.

Neglecting overhead from sequential program sections, SIMD speed-up should be roughly the ratio between the SIMD register size and the relevant data type size assuming a constant clock speed. For x86-64 architectures, the current standard for SIMD instruction sets is SSE4 and its successor AVX which implement SIMD registers between 128 and 512 bits. Working with single (double) precision floating point types, the computational speed-up from SIMD instructions is $4-16$ ($2-8$). The factor two in compute time between single and double precision is relevant for helicity amplitudes, which is considered further in \cref{sec:precision}.

A well-utilised HPC-grade GPU is expected to provide a more significant speed-up than a SIMD CPU for the same problem (again, neglecting all overhead), but there are several reasons to consider also SIMD CPUs for data-parallel programs:
\begin{itemize}
    \item \textbf{Design similarity.} Although there are differences between SIMD CPU and SIMT GPU code, the design philosophy and much of the practical implementation coincides. For an embarrassingly parallel program naturally suited for lockstep processing, the programming overhead in implementing both with respect to just one is not exceptional.
    \item \textbf{Floating-point precision.} For GPUs, hardware-level FP64 instructions are limited to HPC-grade machines. On the other hand, all modern CPUs support higher-precision floating-point operations, including SIMD instructions.
    \item \textbf{Availability.} While HPC-grade GPUs are major investments with limited access in most computing centres, CPUs are necessarily part of every available node. Furthermore, nowadays all major CPUs include SIMD instructions --- for instance, the vast majority of experiment jobs on the WLCG run on nodes supporting AVX2 or later SIMD instructions \cite{Sciaba_2022}. 
\end{itemize}

Clearly SIMD CPUs could provide significant hardware acceleration without the caveats of off-CPU devices such as GPUs. However, they come with a significant restriction of their own: memory layout and management. Since SIMD instructions are applied to sets of adjacent data (see \cref{fig:sisd_vs_simd}), relevant data must be aligned in the sense that the equivalent data between different iterations need to neighbour each other in memory. E.g.\ moving from a scalar operation on the scalar type \texttt{data[1]} to a SIMD operation on the vector type \texttt{data[n]}, elements of \texttt{data} need to be adjacent\footnote{For this minimal example, this change is trivial. In the real world, ensuring adjacency becomes more difficult, especially when vectorising existing (multi-dimensional) arrays.}. Additionally, since the same instruction is applied across the entire SIMD register, for an $n-$fold vector register it is beneficial for the $n$ vectorised iterations follow identical program paths, lest an erroneous instruction be applied to part of the register.
 
\subsection{Single instruction, multiple threads}
\label{sec:simt}

\begin{figure}[t]
    \centering
    \includegraphics[height=3.2cm]{figs/ZW/SISD.pdf}
    \hspace{1.4cm}
    \includegraphics[height=3.2cm]{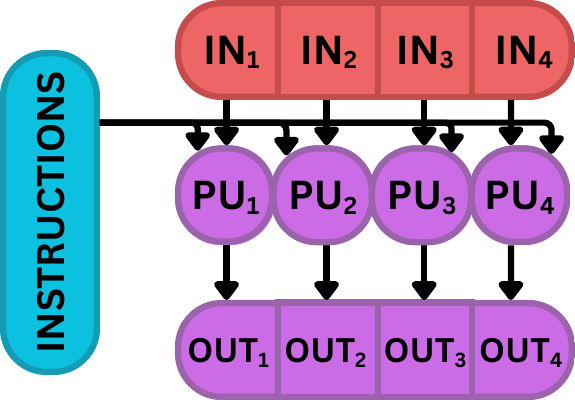}
    \caption{Simple illustration of the difference between SISD (left) and SIMT (right) processing. In SISD, a single PU is loaded with a single datum and a single instruction which outputs a single modified datum. In SIMT, several PUs are acting at once in lockstep and each one can be considered ``identical'' to a single SISD PU, but the instruction fetch is shared between the PUs such that each PU is loaded with the exact same instruction.}
    \label{fig:sisd_vs_simt}
\end{figure}

The SIMD method of instruction-level data parallelism is limited by register size and available instruction sets. Another, more scalable method is to ``simply'' increase the number of processing units, as is the case for SIMT machines such as GPUs. By restricting PUs to smaller instruction sets and shared instruction fetches, the amount of threads that can be produced and run on a single machine increases dramatically. Each thread has its own register, but they act synchronously in lockstep. This is graphically represented in \cref{fig:sisd_vs_simt}.

This design comes with its own strengths and weaknesses. Unlike SIMD CPUs, where synchronicity can be destructive at program branches, threads on a SIMT GPU act synchronously but independently, meaning threads will simply idle at failed conditionals rather than destroy the data. However, the scale of parallelism in a GPU is much greater than SIMD registers, necessitating far larger problem sizes to make optimal use of the machine and making substantial branching in a program wasteful. Furthermore, the singular data entry point of an off-CPU device makes memory management a key consideration for SIMT programming --- latency in device communication can severely limit any possible speed-up. 

The former issue is dealt with at the hardware-level by grouping PUs together in separate sets, limiting the extent of lockstep to these sets. Using Nvidia's nomenclature, these are called \textit{warps} and on Nvidia (AMD) devices consist of 32 (64) PUs with a shared instruction fetch. At the software level, though, work is defined in terms of \textit{blocks}, sets of threads defined at runtime which are scheduled to run the same code. To make full use of the SIMT GPU, it is immediately clear that block size must be a multiple of the warp size.

As for the latter issue, it must be dealt with on the software side. GPU memory hierarchy is structurally similar to that of a CPU, with the caveat that larger memory pools are shared between greater sets of PUs: each thread has its own cache (\textit{local memory}); each block has a joint memory pool (\textit{shared memory}); and the full device has both a \textit{global cache} and \textit{global memory}. However, there is a far greater bottleneck than on-device latency: host-device latency. Regardless of bandwidth, the fact that the CPU and GPU are separate machines makes memory transfer slow. The main consideration when designing SIMT GPU code should consequently be maximising continuous on-device code execution and minimising instances of host-device communication. Of course, the on-device memory hierarchy is still an important consideration which must be kept in mind; threads in a warp execute instructions in lockstep, including the load instruction. So-called coalesced memory access can be performed when each thread in a warp loads data from consecutive memory addresses; failing to properly set up data for coalesced access means each access must be scheduled independently of one another.

An additional concern for GPU programming already mentioned in \cref{sec:simd} is data type availability. As native FP64 operations are only supported on HPC-grade GPUs, programs requiring high-precision floating point calculations are unlikely to utilise more generally available lower-tier GPUs well. While FP32 (or even integer types) is sufficient for many problems, helicity amplitudes in standard gauge choices involve large cancellations between different Feynman diagrams risking catastrophic cancellations; this topic is elaborated on in \cref{sec:precision}.

\section{Event-level data parallelism}
\label{sec:datapara_evgen}

While large swaths of HEP software includes significant branching, this is not the case for scattering amplitudes. A scattering amplitude is, somewhat oversimplified, the probability of a particle-level process occurring for a specific momentum distribution within a given QFT. The cross section of a process is the integral of the (absolute squared) scattering amplitude, and given the stochastic nature of quantum mechanics the phase space of available momentum distributions should be sampled according to the underlying scattering amplitudes to make statistically unbiased empirical predictions about experimental observables. Particularly, while the integrated cross section generally does not have closed-form solution, the scattering amplitude (at any given perturbative order) does, and at leading order this formula is identical across the entirety of phase space.\footnote{Note, however, that phase-space \textit{handling} is not identical over the full phase space, which can create branching. How this is handled in \cudacpp{} is specified in \cref{sec:technicalevgen}}. In short, for LO event generation the exact same (largely branch-free) calculation is done for a substantial amount of input momenta, making it an ideal candidate for data parallelism.

The structure of helicity amplitudes as used by \mg{} is provided in \cref{sec:technicalstandalone} alongside technical details of the presented data-parallel implementation in \cudacpp{}. This implementation is written in C++, with compile-time \texttt{typedef} statements masking operations for either SIMD C++ or SIMT GPU backend compilations. To avoid rewriting the entire \mg{} (LO) event generator \textsc{Mad\-Event} in C++, though, the driving event generator largely maintains the original Fortran codebase albeit with a multi-event interface as well as a Fortran/C++ program interface for communication between the two different codes. This interfacing, alongside the algorithmic changes made to event generation, is detailed in \cref{sec:technicalevgen}. Then, in \cref{sec:gridpacks}, a brief overview of the Python driver for \textsc{MadEvent} is given, alongside a description of pre-compiled gridpacks which are typically used by experimental collaborations and how these have been modified to better utilise hardware with \cudacpp{}.

\subsection{Parallel scattering amplitudes}
\label{sec:technicalstandalone}

Helicity amplitudes are a Feynman diagram method for evaluating squared amplitudes where --- for each parton-level process --- the squared scattering amplitudes for all (non-vanishing) spin configurations are calculated independently and summed over. In \mg{} this is done by evaluating the explicit wavefunction form of external particles and propagators until all remaining particles enter a final vertex where relevant inner products are taken.

As an example, consider the minimal process $g\,g\to t\,\overline{t}$; first, the external particle wavefunctions are evaluated, which in the \mg{}-generated Fortran code for this process is done by:
\begin{lstlisting}[language=madgraph]
!     External partons
      CALL VXXXXX(P(0,1),ZERO,NHEL(1),-1,W(1,1))
      CALL VXXXXX(P(0,2),ZERO,NHEL(2),-1,W(1,2))
      CALL OXXXXX(P(0,3),MDL_MT,NHEL(3),+1,W(1,3))
      CALL IXXXXX(P(0,4),MDL_MT,NHEL(4),-1,W(1,4))
\end{lstlisting}
where the \texttt{VXXXXX} routines evaluate the initial state gluon wavefunctions from the corresponding physics parameters (momenta, mass, etc.) and output them into the corresponding dimension of the array of wavefunctions \texttt{W}, and the \texttt{O}/\texttt{IXXXXX} routines do the same for the top-antitop quark pair. The corresponding snippet in \cudacpp{} would be
\begin{lstlisting}[language=cudacpp]
// External partons
vxxxxx<M_ACCESS,W_ACCESS>(momenta,0.,cHel[ihel][0],-1,w_fp[0],0);
vxxxxx<M_ACCESS,W_ACCESS>(momenta,0.,cHel[ihel][1],-1,w_fp[1],1);
oxxxxx<M_ACCESS,W_ACCESS>(momenta,cIPD[0],cHel[ihel][2],+1,w_fp[2],2);
ixxxxx<M_ACCESS,W_ACCESS>(momenta,cIPD[0],cHel[ihel][3],-1,w_fp[3],3);
\end{lstlisting}
where the only significant difference in structure is the inclusion of templated memory access classes in the declaration of the wavefunctions. When compiled for SIMD CPUs, the momenta and wavefunction arrays will have an additional dimension corresponding to the vectorisation and each call to these functions is actually a loop over the elements of this dimension. On a SIMT GPU, on the other hand, these arrays do still have an extra dimension in memory, but each function call only evaluates the routines for the specific entries corresponding to the given thread.

Next, two (or more) external partons are merged into an off-shell propagator. Taking e.g. the gluon-mediated $s$-channel, the relevant Fortran code is
\begin{lstlisting}[language=madgraph,firstnumber=6]
!     Wavefunction(s) for diagram number 1
      CALL VVV1P0_1(W(1,1),W(1,2),GC_10,ZERO,ZERO,W(1,5))
\end{lstlisting}
and the corresponding \cudacpp{} code is
\begin{lstlisting}[language=cudacpp,firstnumber=6]
// *** DIAGRAM 1 OF 3 ***
// Wavefunction(s) for diagram number 1
VVV1P0_1<W_ACCESS,CD_ACCESS>(
    w_fp[0],w_fp[1],COUPs[0],1.0,0.,0.,w_fp[4]
);
\end{lstlisting}
which, again, is almost identical to the Fortran code. The significant change is in the handling of the vertex' coupling constant: in Fortran, this is passed as a reference to the scalar \texttt{GC\_10}, while in \cudacpp{} it is passed as a pointer to an element of an array \texttt{COUPs[0]}. An additional argument is also included, a floating point $\pm1$, corresponding to the sign of the coupling constant --- in Fortran, this would be given as a multiplication in the function call, but as \cudacpp{} passes around pointers rather than references this is impossible.

Finally, the remaining three (or more) particles are combined in a final vertex which performs the relevant inner products to map these wavefunctions to a complex scalar, which in Fortran is given by
\begin{lstlisting}[language=madgraph,firstnumber=8]
!     Amplitude(s) for diagram number 1
      CALL FFV1_0(W(1,4),W(1,3),W(1,5),GC_11,AMP(1))
\end{lstlisting}
and in \cudacpp{} is written as
\begin{lstlisting}[language=cudacpp,firstnumber=11]
// Amplitude(s) for diagram number 1
FFV1_0<W_ACCESS,A_ACCESS,CD_ACCESS>(
    w_fp[3],w_fp[2],w_fp[4],COUPs[1],1.0,&amp_fp[0]
);
\end{lstlisting}
which have the same form as the previous vertex functions but now return the helicity amplitude, as denoted by the \texttt{\_0} in the name of the function. The last two steps are repeated with the corresponding routines for each individual diagram, after which the colour algebraic inner product is taken to determine the absolute squared amplitude. Note that this sequence is identical across all of phase space and that the only branching occurs in the helicity states of the external legs. Since \mg{} by default performs helicity summation rather than sampling, these branches are looped over in a lockstep-friendly manner\footnote{Although there are no plans to implement helicity sampling in \cudacpp{}, doing so in a lockstep-friendly manner would not be difficult. It would require either sample bunching (such that $n$ phase-space points use the same helicity configurations) or rewriting external wavefunctions with branchless arithmetic; the former is simple and unlikely to create a noteworthy bias, and the latter would entail few and straightforward code modifications.}.

While the \cudacpp{} code structure is a one-to-one port of the corresponding Fortran code, certain considerations are hidden in the function call structure above. Most significantly all floating point numbers have been replaced by a generic \texttt{FPTYPE}, defined at compile-time to be FP64 or FP32 and scalar or SIMD types, and the corresponding numerical operators (i.e. $+$, $*$, etc.) have been overloaded to compile to the corresponding scalar or SIMD ones. A generic \texttt{Kernel} class internally handles function calls and memory access, such that the same scattering amplitude calls in an executable can be used for local SIMD CPU evaluations or on-device SIMT GPU evaluations. These compile-time aliases and definitions are generic; the same \cudacpp{} program can be compiled not only to C++ for SIMD CPUs and CUDA for Nvidia GPUs, but also to HIP for AMD GPUs using minimal internal API aliases within \cudacpp{}. 

\subsection{Event generation interface}
\label{sec:technicalevgen}

While helicity amplitudes are well-suited for data parallelism, \textsc{MadEvent} was written with the assumption of sequential execution. Although amplitudes are handled within  \cudacpp{}, development has also entailed alterations to the interface between event generation and amplitudes which previously assumed scalar return values. This multi-event interface is officially included in \mg{} as of version 3.6.0, released on September 30th, 2024.

The multi-event interface is a small abstraction layer between the higher-level event generation routines and squared amplitude evaluations, primarily entailing the appendage of a dimension to all event-specific data used in amplitudes. The size of this dimension is given by the compile-time integer \vecsize{}, i.e. the ``degree of vectorisation'' meaning how many squared amplitudes should be evaluated per call to the amplitude routines. By default \vecsize{} is set to 1 (reducing to the original scalar algorithm of \mg{} prior to 3.6.x), but it can be set arbitrarily. For \vecsize{} greater than 1 this allows for multi-threading over events in native \mg{} event generation, and allows for SIMD and SIMT parallelism using \cudacpp{}.

In addition to the vectorisation of event-specific data, the multi-event interface has seen overhauls in data access. Due to the previous scalar nature of \mg{} event generation, the codebase makes extensive use of stateful routines and public data access using common blocks and static variables. To ensure data locality and safety, such data have been made into arguments for amplitude routines rather than global variables. This modification ensures that separate calls to the amplitude routines will act on distinct data, without having to keep track of indexing at the level of the individual amplitudes.

On the \cudacpp{} side, a \texttt{Bridge} class handles internal memory management (such as transposition between column- and row-major layout) and has a functionality wrapper called \texttt{fBridge}. A Fortran interface for \texttt{fBridge} then enables access to \cudacpp{} functionality from the \textsc{MadEvent} multi-event interface. This interface is minimal, in the sense that it wraps all necessary functionality for event generation without providing access to any of the internal structures or additional functionality; from the Fortran executables, \texttt{Bridge} objects can only be constructed, executed, and destroyed. 

One final consideration is phase-space handling: while the helicity amplitude routines for individual parton configurations are non-branching, the upstream \mg{} project by default groups together parton configurations with sufficiently similar kinematic structures, unifying them into a single integration channel --- and, consequently, compiled executable. By default, \mg{} performs MC over symmetric integration channels and external quark flavours\footnote{Technically, this same treatment is used for initial state symmetry (e.g. $u \,d \to \cdots$ and $d \, u \to \cdots$) but the full computation is then identical under an index change $1 \leftrightarrow2$.}, creating branching paths in the integration despite the perfect lockstep processing in the individual amplitude calls.

Considering the MC over phase-space symmetry, the idea is to have a single integration grid for phase-space mappings that are identical up to the ordering of final state particles. This reduces the necessary amount of training iterations to determine the phase-space grid up to any given precision (and the resulting amount of individual grids stored to disk); however, the differing multi-channel weights associated with the single diagram enhancement method used in \textsc{MadEvent} \cite{Maltoni:2002qb} makes the resulting computations non-identical. Given that diagram enhancement is a central aspect of \textsc{MadEvent} event generation this branching is maintained in \cudacpp{}, but in order to maintain perfect lockstep over phase-space points the MC algorithm now occurs over blocks of $n$ events rather than single ones. To avoid biasing the results, $n$ must be kept much smaller than the number of evaluated phase-space points; $n$ should be kept as small as possible, ideally at the register or warp size of the corresponding SIMD CPU or SIMT GPU. The value of the block size $n$ is configurable by the user, as detailed in \cref{appendix:manual}.

For the MC over external parton flavours, \mg{} generates distinct Fortran code for each individual parton configuration (save for instances of perfectly identical helicity amplitudes) and MC is performed at the level of the individual calls to amplitude routines. This technique could also have been applied for \cudacpp{} using the same block grouping method as for parton ordering, but it would bring some complications: first and foremost, the resulting kernel size would blow up, wasting GPU memory especially for very complex processes; secondly, while MC over parton flavours reduces \textit{total} code size and grid training requirements, it does not speed up the event generation algorithm itself. However, it is worth noting that this choice greatly increases \cudacpp{} compilation time; motivated partially by this fact, there is ongoing work on restructuring flavour handling within \mg{}. Instead of defining flavours as distinct external partons, they are planned to be treated akin to helicity (i.e. as a quantum number to sum or sample over) which moves the flavour handling into the scattering amplitude itself and minimises code size both at compile- and runtime.

\subsection{Event generation and gridpacks}
\label{sec:gridpacks}

\Cref{sec:technicalstandalone,sec:technicalevgen} illustrate the structure of the compiled executables used for event generation in \mg{}, but do not detail the full runtime structure. As process structure is encoded into the \mg{}-generated code, each distinct parton configuration contributing to a given process is compiled into a separate \textsc{MadEvent} executable; scheduling and launching these executables and combining their output into a single LHE file is done by a \textsc{MadEvent} driver written in Python (henceforth called the driver).

As a high-level description, the driver can be viewed as a set of workers and jobs. At runtime, a number of workers are launched, and each \textsc{MadEvent} executable has its corresponding phase space divided, defining each region assigned to each executable as a job. The jobs are put into a queue accessible to the workers, which pull the next available job from the queue, handle the runtime arguments, and launch it. When a job is finished, the relevant worker pulls the next one from the queue, until eventually all jobs are finished.

Due to its model-generic structure and being a code generator, event generation using \mg{} comes with significant overhead: the code needs to be generated and compiled before even estimating the underlying phase-space distribution, which must be fixed up to the desired MC precision of the final unweighted sample. To avoid having to repeat this process each time events are generated, \mg{} can generate so-called \textit{gridpacks}: minimal pre-compiled executables for a given process with pre-defined phase-space separations based on the corresponding phase-space distribution, as well as a minimal version of the Python driver.

Prior to this work, which is included in the official version of \mg{} as of version 3.6.2, \mg{} gridpacks only took two necessary and one optional runtime argument: \texttt{nb\_events}, the number of unweighted events requested; \texttt{seed}, the seed used to initialise the random generation of phase-space points; and \texttt{granularity} (optional), the minimum number of weighted events to generate before unweighting any given integration channel.

This gridpack setup had two major restrictions based on assumptions on use cases: only one worker was ever launched (limiting gridpack runs to single CPU cores), and each job could generate a maximum of 2 500 unweighted events. The former restriction was made assuming it would generally be better to schedule $n$ single-core executions than one $n$-core execution in a distributed environment, and the latter is a trick to linearise the problem and avoid any potential scaling in quadrature of the problem at hand. 2 500 events are sufficiently many to ensure that any overhead from launching \textsc{MadEvent} executables is negligible compared to \textsc{MadEvent} runtimes. If more than 2 500 events are needed for a given integration channel, two separate jobs are scheduled to reach the number of required events asked by the user.

Considering GPU offloading, these restrictions become more problematic: using a single core to launch GPU kernels makes it practically impossible to fully utilise HPC-grade GPUs, especially for processes with many different integration channels; on the other hand, given a realistic unweighting efficiency and very varied process complexities, 2 500 unweighted events can have significantly varying runtimes between integration channels, resulting in singular jobs taking up a vast majority of runtimes when they could have been split up among workers.

Consequently, two additional optional runtime flags have been added to \mg{} gridpacks: \texttt{-p} (\texttt{-{}-parallel}), setting the number of workers available in the driver and consequently the number of simultaneous scheduled executions (which is not necessarily the same as number of simultaneous executions, if \texttt{-p} exceeds the number of available cores); and \texttt{-m} (\texttt{-{}-maxevts}), setting the maximal number of unweighted events generated in each job. These flags are, as stated, optional and will take on the default values of 1 and 2 500 respectively, reducing to the parameters previously hardcoded.

\section{Standalone application}
\label{sec:standalone}

The \texttt{standalone} output in \mg{} (henceforth just standalone) is a minimal executable for evaluating summed squared helicity amplitudes, foregoing all of the surrounding infrastructure of e.g. event generation. When run, it generates a random phase-space point for the given parton-level process --- using the \texttt{RAMBO} algorithm \cite{Rambo} --- and evaluates the spin-summed scattering amplitude squared at that point. While standalone has limited relevance on its own, it serves as an encapsulation of \mg{}-generated scattering amplitudes and can be used as a testbed both for the accuracy and relevance of any changes made to amplitude evaluations. Since the work on \cudacpp{} primarily concerns the data-parallel implementation of these, standalone can be used to verify physics accuracy for the amplitudes as well as provide an upper bound for the acceleration \cudacpp{} can provide in real-world use cases.

In \cref{sec:precision}, analyses of the numerical stability of summed squared helicity amplitudes are provided using the round-off error estimation library CADNA \cite{Eberhart_REC_2015,FJ_CADNA_2015} as well as physics-motivated numerical tests, and while the results suggest there may be issues in evaluating helicity amplitudes entirely using FP32 they also illustrate aspects of the calculation that may not require FP64 precision. Following this study on numerical precision, \cref{sec:standalone_simd,sec:standalone_gpu} showcase \cudacpp{} standalone acceleration for SIMD CPUs and SIMT GPUs respectively.

\subsection{Numerical stability}
\label{sec:precision}

\begin{table}[t]
    \centering
    \begin{tabular}{|c|c|c|c|} \hline
         FP type & \textbf{Worst} & \textbf{Best} & \textbf{Mean} \\ \hline
         \textbf{FP32} & 1 & 6 & 4.30 \\ \hline
        \textbf{FP64} & 10 & 14 & 12.96 \\ \hline
        \textbf{mixed} & 5 & 8 & 6.21 \\ \hline
    \end{tabular}
    \caption{Number of accurate significant digits in squared scattering amplitudes for the process $g g \to t \overline{t} + 3g$ evaluated using FP32, FP64, and the physically motivated mixed mode, as measured using CADNA. Each measurement consists of 8000 phase-space points. In mixed precision, helicity amplitudes are evaluated in FP64 to properly capture Feynman diagram interference, but are cast to FP32 before evaluation of the colour algebra. Mixed precision is the default mode in \cudacpp{}. For comparison, theoretical uncertainty in leading order calculations is typically estimated to be on the order of 30\% to 60\%.}
    \label{tab:cadnaresults}
\end{table}

Using standard gauge choices in standard representations, Feynman diagram methods such as helicity amplitudes involve large cancellations between different contributions. Consequently, it is generally assumed that helicity amplitudes require FP64 precision to avoid catastrophic cancellations despite the sizeable theoretical uncertainty of leading order calculations. However, these cancellations occur specifically in the kinematics of the process; for complex QCD processes, which constitute the vast majority of physics at the LHC, a significant fraction of compute is used evaluating colour matrix algebra. Unlike the kinematic helicity amplitudes, the colour algebra constitutes an inner product of colour-ordered scattering amplitudes and a process-dependent colour matrix of integer coefficients \cite{Lifson:2022mxf}, suggesting ``mixed precision'' calculations where helicity amplitudes are evaluated with FP64 before being cast to FP32 when evaluating the colour algebra is likely to be numerically stable. The speed-up provided by this mixed precision mode is detailed in \cref{sec:standalone_gpu}.

The accuracy of scattering amplitudes squared evaluated using different levels of precision was tested using CADNA. CADNA is a library for estimating round-off error propagation in software using Discrete Stochastic Arithmetic by replacing numerical types with corresponding CADNA ones, which runs a given program multiple times with random rounding modes to probabilistically measure the number of accurate significant digits as compared to exact calculations. A summary of CADNA results for the 7-body process $g g \to t \overline{t} \, g  g g$ is given in \cref{tab:cadnaresults}, where 8000 randomly generated phase-space points were tested for each precision mode.

\begin{figure}[t!]
    \centering
    \includegraphics[width=0.95\linewidth]{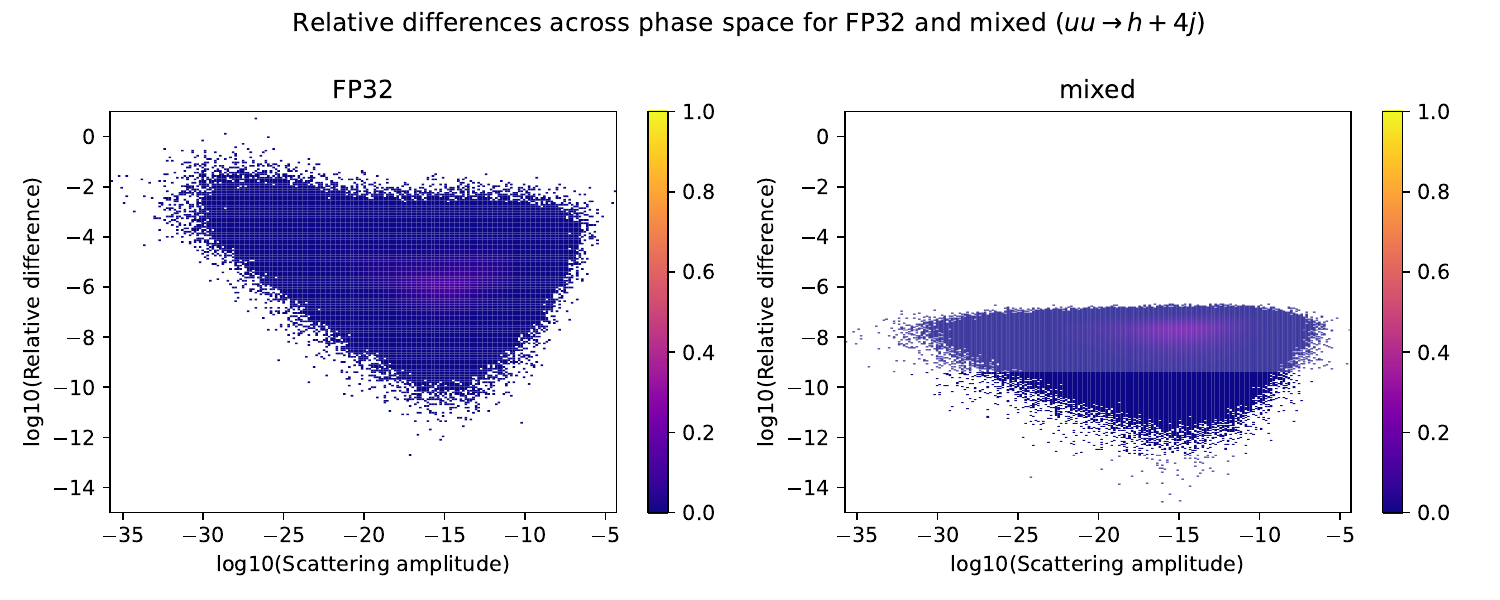}
    \caption{Relative difference in scattering amplitudes squared evaluated in FP32 (left) and mixed precision (right) compared to the amplitude evaluated in FP64 plotted against the magnitude of the squared amplitude evaluated in FP64, for the VBF process $u u \to h \, +4j$ with $j$ any massless QCD jet. Mixed precision refers to evaluating helicity amplitudes in FP64 but colour algebra in FP32. The colour map denotes fraction of events in each of the 200 bins. Each plot consists of roughly eight million identical phase-space points generated with pre-defined phase-space grids from \textsc{MadEvent}. Note that the relative difference refers to the first differing digit, such that accuracy is one order of magnitude larger than plotted. Additionally, FP32 corresponds to roughly seven significant decimal digits; any smaller error is coincidental.}
    \label{fig:vbfprecision}
\end{figure}

While CADNA provides well-founded and justified measurements of numerical precision, there are some issues with generalising such tests in an environment like \mg{}. Long execution times make substantial tests of e.g. millions of phase-space points impractical, especially when considering the significant overhaul necessary to implement these tests in a code generator like \mg{}. As such, further tests with large samples have been performed by evaluating identical phase-space points across the three precision modes, taking evaluations in FP64 as a baseline. The precision of FP32 and mixed FP64/FP32 modes are then estimated by their difference from this baseline.

A plethora of processes were tested in this manner, although a physically motivated representative ``worst case scenario'' is given by vector boson fusion processes (VBF) which generally involve the most significant cancellations between different Feynman diagrams of any leading-order SM processes, which in this case are related to gauge cancellation. A two-dimensional histogram of the spread of this error is given in \cref{fig:vbfprecision}, with squared scattering amplitude magnitudes (as evaluated in FP64) along the $x$-axes and absolute relative errors (given as the ratio with the FP64 amplitude squared minus one) along the $y$-axes. The colour map shows what fraction of all roughly eight million events are contained in a given bin. Note that a relative difference on the order of $10^{-n}$ corresponds to an accuracy of $10^{-(n-1)}$, and furthermore that FP32 constitutes approximately seven significant decimal digits, implying any relative errors below $\sim 10^{-8}$ are coincidental cancellations when casting the squared amplitude back to FP64.

\begin{figure}[t!]
    \centering
    \includegraphics[width=0.95\linewidth]{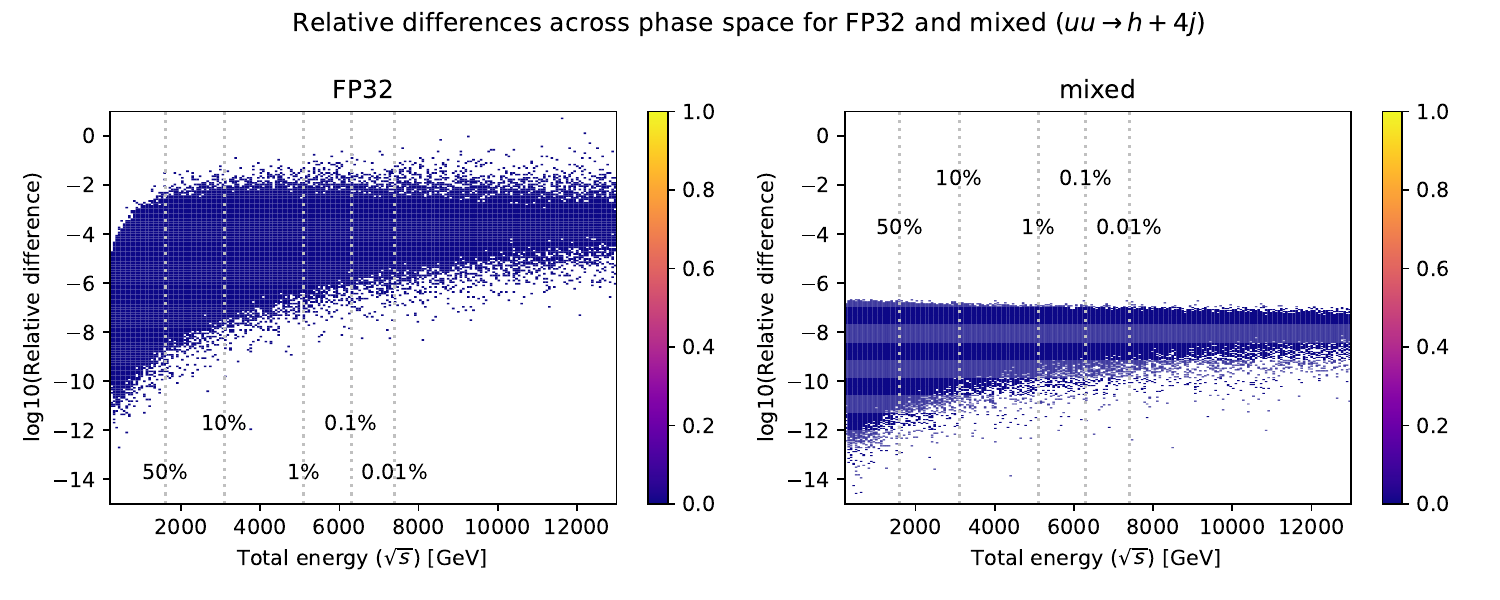}
    \caption{Relative difference in scattering amplitudes squared evaluated in FP32 (left) and mixed precision (right) compared to the amplitude evaluated in FP64 plotted against the total energy $\sqrt{\hat{s}}$ of the interaction, for the VBF process $u u \to h \, +4j$ with $j$ any massless QCD jet. The colour map denotes fraction of events in each of the 200 bins. Each plot consists of roughly eight million identical phase-space points generated with pre-defined phase-space grids from \textsc{MadEvent}. Dotted lines are labelled with the percentage of the total cross section of this process at LHC energies encapsulated by events to the \textit{right} of the line (i.e. at a higher energy than denoted), or equivalently the fraction of unweighted events expected to be sampled at energies higher than said line.}
    \label{fig:vbfenergy}
\end{figure}

The right-hand plot in \cref{fig:vbfprecision} illustrates that mixed precision worsens the numerical results little more than the loss in casting amplitudes from FP64 to FP32. This coincides with the results in \cref{tab:cadnaresults} (although colour algebra constitutes a significantly smaller part of VBF than pure QCD processes), justifying the decision to make mixed precision the default in \cudacpp{}; however, any speed-up is limited to the colour algebra, and the need for FP64 remains. The left-hand plot poses less obvious considerations. Clearly, rounding errors sometimes exceed LO theoretical precision ($\sim 30-60\%$). However, while FP32 rounding errors may coincide with or exceed theoretical uncertainty, there is no a priori reason to assume they will bias generated samples; in fact, some small non-conclusive tests suggest rounding errors will ``wash out'' and have little to no effect on statistics. 

As an additional test, the same phase-space points as above are plotted against the total energy of the corresponding interaction ($\sqrt{\hat{s}}$) in \cref{fig:vbfenergy} to determine whether these errors are exclusive to irrelevant regions of phase space. In \cref{fig:vbfenergy} dotted lines have been added denoting what fraction of the total cross section is contributed by regions of the phase space with energy higher than these lines. This test highlights that energies upwards of $5\,300$ GeV are still significant for this process, where a notable fraction of events have a relative difference of $10^{-1}$, i.e. only one correct significant digit.

These points are insufficient to determine whether FP32 would suffice for helicity amplitudes in standard gauge choices\footnote{There has been recent development in the Feynman diagram gauge, an alternative gauge choice which minimises Feynman diagram interference, reducing the differences between individual amplitudes and the resulting squared scattering amplitudes \cite{Hagiwara:2020tbx,Chen:2022gxv,Chen:2022xlg,Hagiwara:2024xdh}. This, or other methods that minimise gauge cancellations, could potentially make calculations sufficiently stable that FP32 would be definitively sufficient.}. More extensive validations are left for future work. At this stage, it can only be said that FP32 computations are likely to be problematic and should not be used without case-by-case validations. Nevertheless, the option is left to the user.

\subsection{SIMD CPU results}
\label{sec:standalone_simd}

\begin{figure}[t!]
    \centering
    \includegraphics[width=0.45\linewidth]{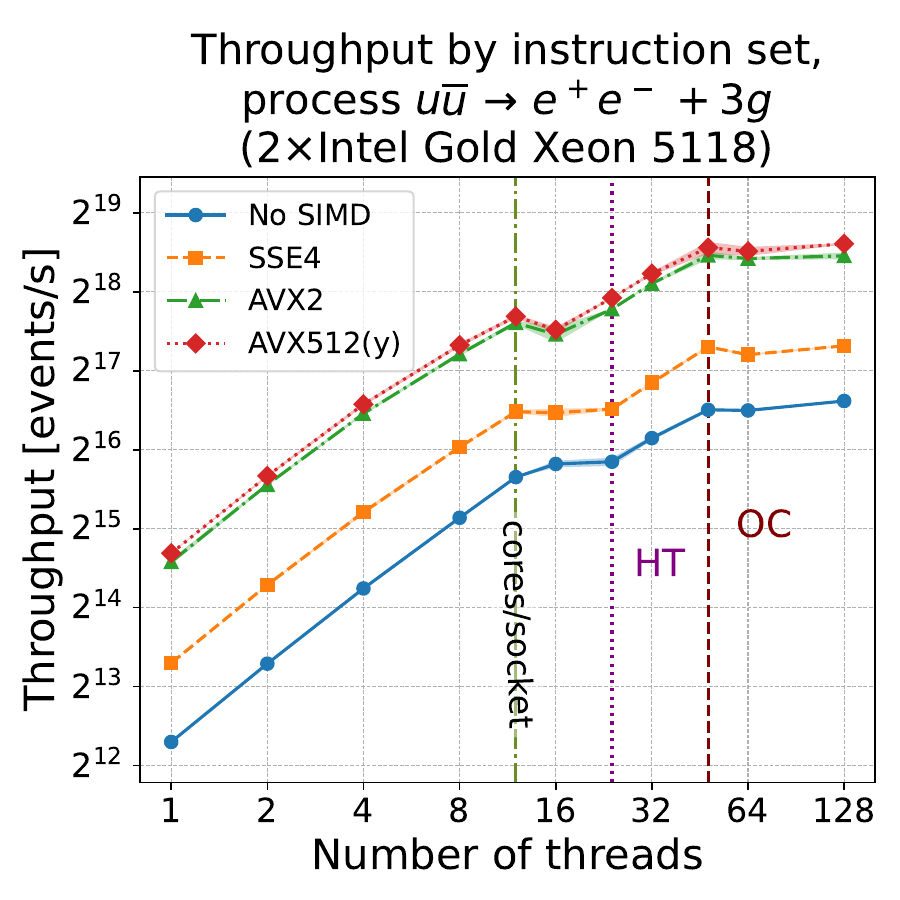}
    \includegraphics[width=0.45\linewidth]{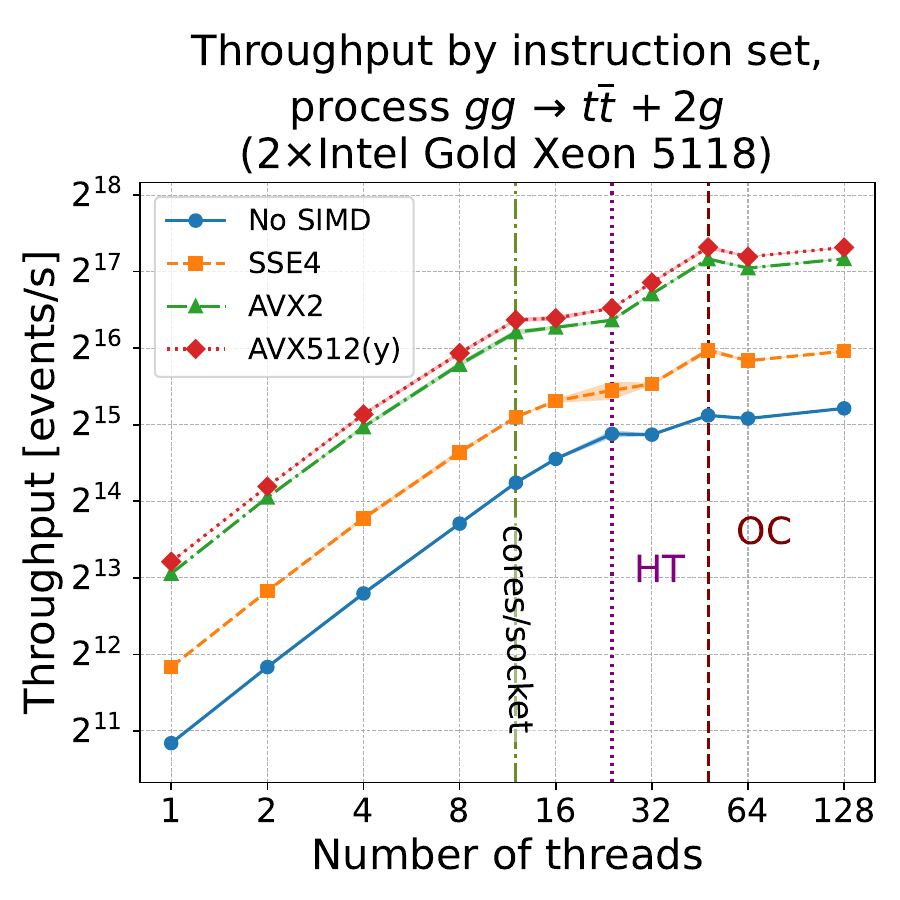}
    \caption{\cudacpp{} standalone throughput (including random number generation, phase-space mapping, and scattering amplitudes) for the SM Drell-Yan process $u \overline{u}\to e^+ e^- +3g$ (left) and QCD process $gg\to t\overline{t}+2g$ (right) for varying SIMD instruction sets, varying the number of threads while running multithreaded. Each of the two CPUs, Intel Gold Xeon 5118, has 12 physical cores and supports hyper-threading for a total of 24 logical units per socket. Regions with hyper-threading (HT) and overcommitment (OC) are marked. Each point is given by the average throughput from 10 \cudacpp{} standalone runs on the CPU with $2^{18}$ phase-space points each, and standard deviations are highlighted albeit typically too small to see.}
    \label{fig:standalone_throughput_skylake}
\end{figure}

\begin{figure}[t!]
    \centering
    \includegraphics[width=0.45\linewidth]{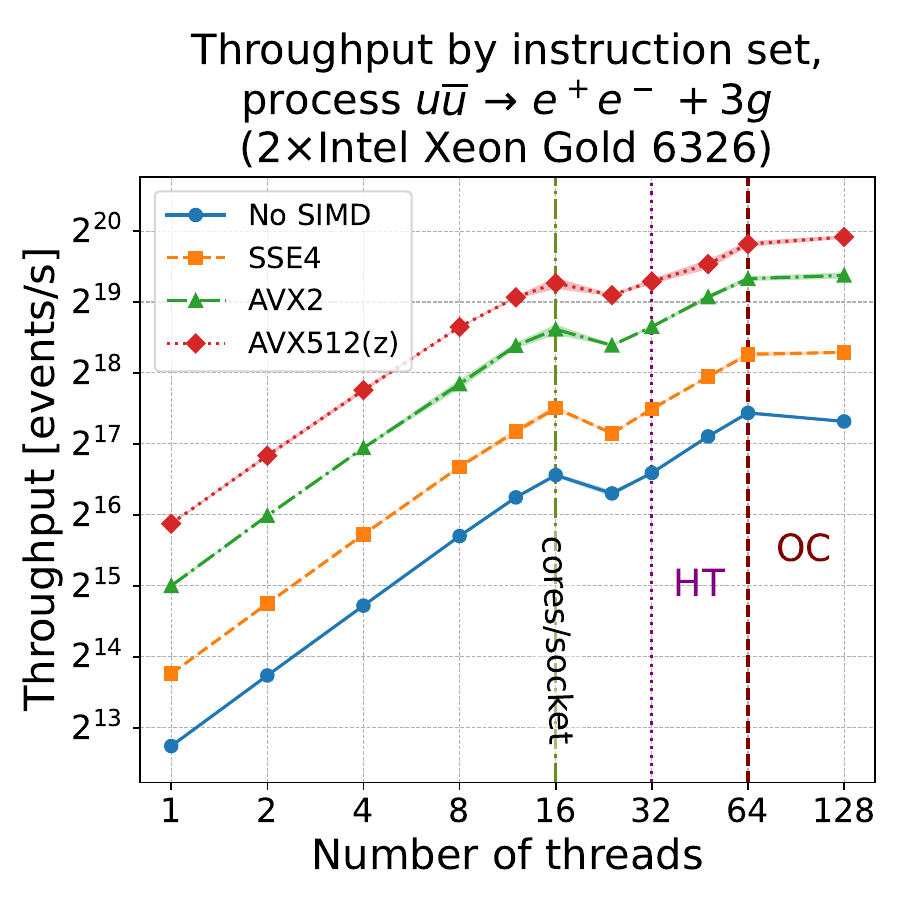}
    \includegraphics[width=0.45\linewidth]{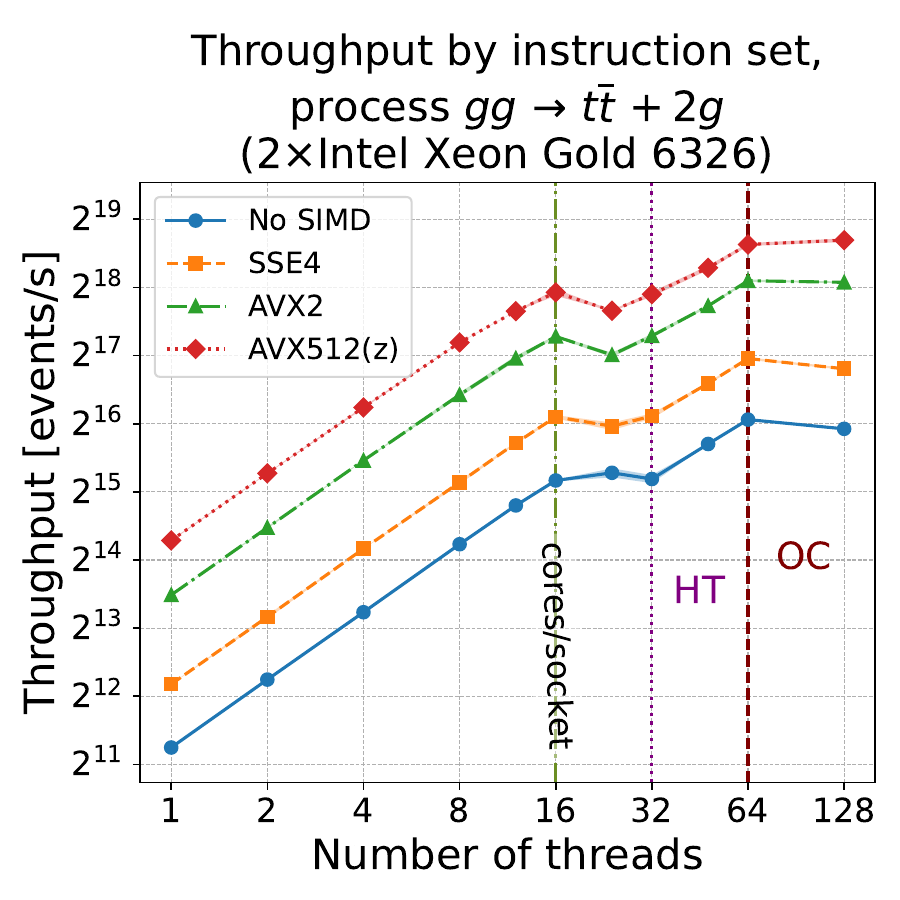}
    \caption{\cudacpp{} standalone throughput (including random number generation, phase-space mapping, and scattering amplitudes) for the SM Drell-Yan process $u \overline{u}\to e^+ e^- +3g$ (left) and QCD process $gg\to t\overline{t}+2g$ (right) for varying SIMD instruction sets, varying the number of threads while running multithreaded. Each of the two CPUs, Intel Gold Xeon 6326, has 16 physical cores and supports hyper-threading for a total of 32 logical units per socket. Regions with hyper-threading (HT) and overcommitment (OC) are marked. Each point is given by the average throughput from 10 \cudacpp{} standalone runs on the node with $2^{18}$ phase-space points each, and standard deviations are highlighted albeit typically too small to see.}
    \label{fig:standalone_throughput_icelake}
\end{figure}

\begin{figure}[t!]
    \centering
    \includegraphics[width=0.45\linewidth]{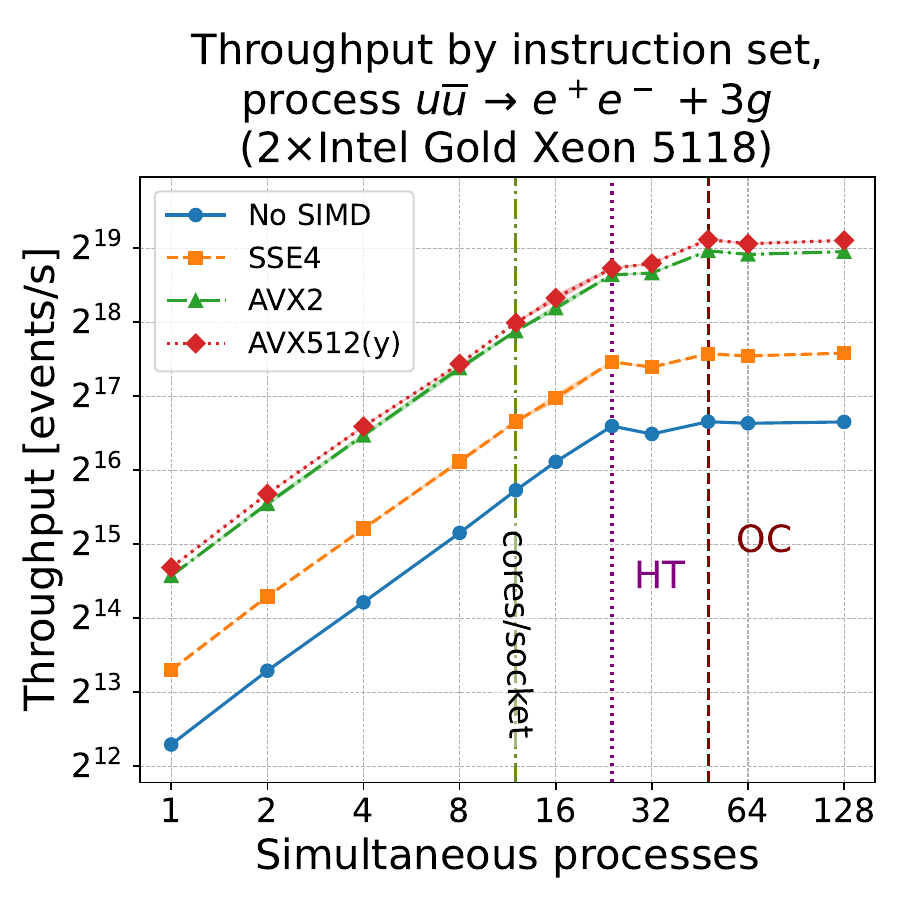}
    \includegraphics[width=0.45\linewidth]{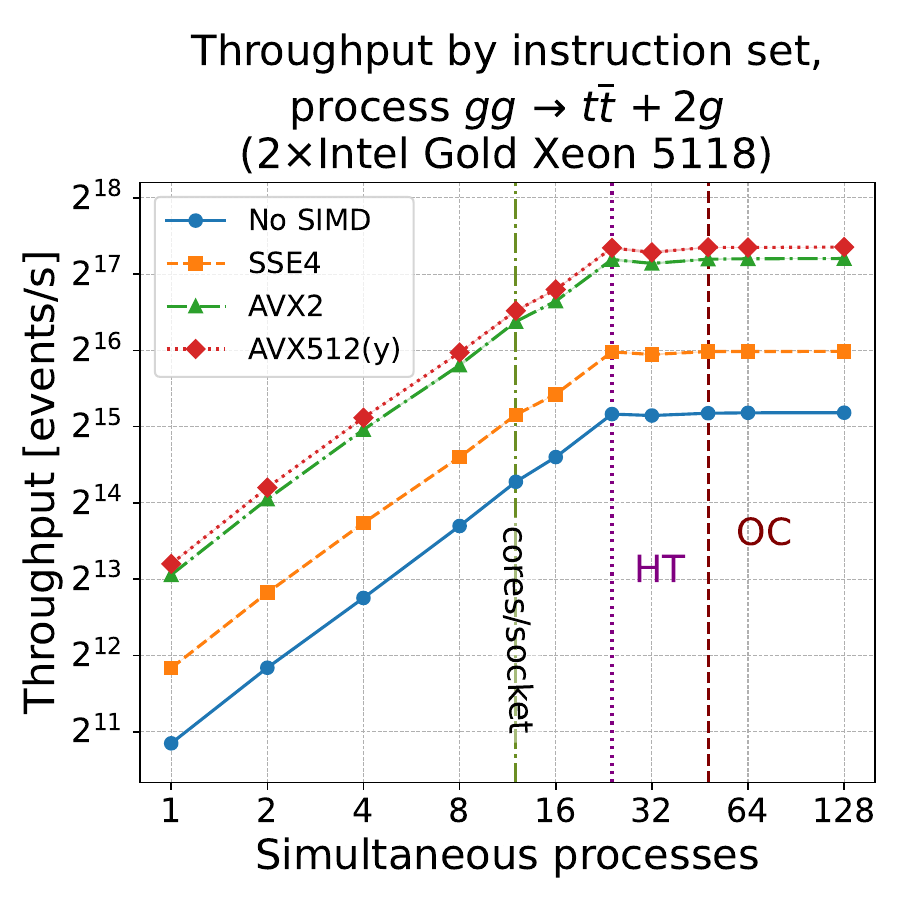}
    \caption{\cudacpp{} standalone throughput (including random number generation, phase-space mapping, and scattering amplitudes) for the SM Drell-Yan process $u \overline{u}\to e^+ e^- \, ggg$ (left) and QCD process $gg\to t\overline{t}\,gg$ (right) for varying SIMD instruction sets, varying the number of simultaneously scheduled executions of the standalone executable. The throughput is defined as the total number of evaluated amplitudes divided by the total real world runtime. Each of the two CPUs, Intel Gold Xeon 5118, has 12 physical cores and supports hyper-threading for a total of 24 logical units per socket. Regions with hyper-threading (HT) and overcommitment (OC) are marked. For each run $n$ standalone executions are simultaneously scheduled to run $2^{18}$ phase-space points each, and each point is given by the average throughput of 10 such runs. Standard deviations are highlighted albeit typically too small to see.}
    \label{fig:standalone_throughput_skylake_mp}
\end{figure}

\begin{figure}[t!]
    \centering
    \includegraphics[width=0.45\linewidth]{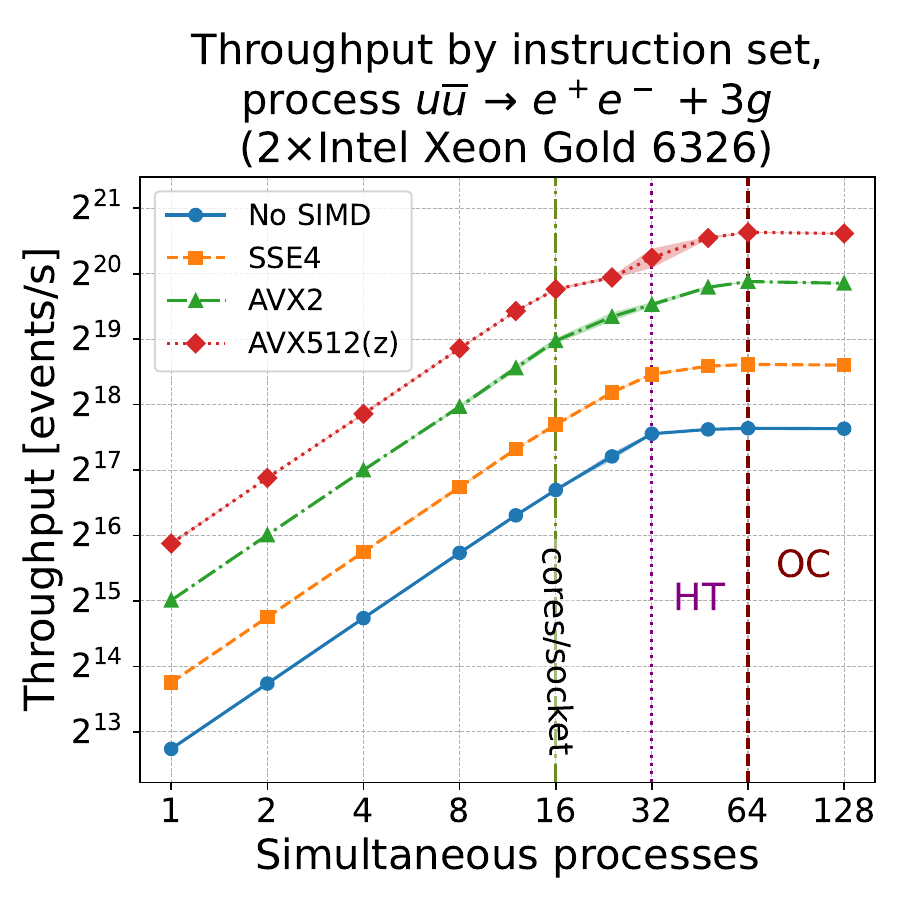}
    \includegraphics[width=0.45\linewidth]{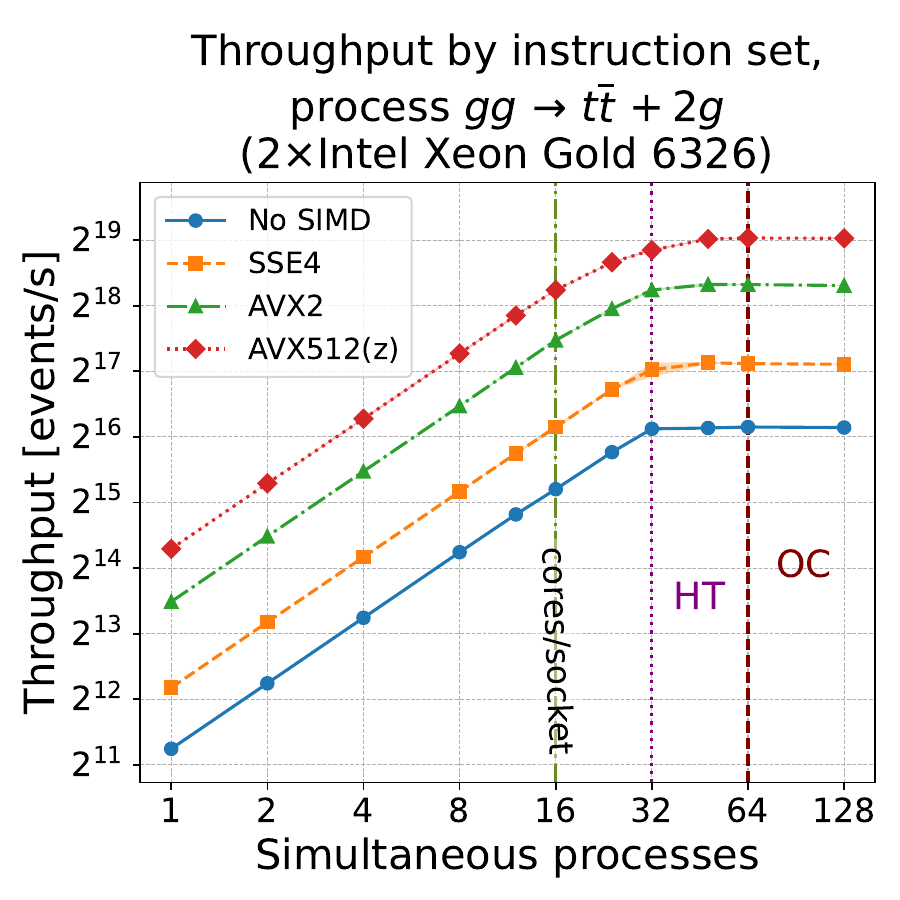}
    \caption{\cudacpp{} standalone throughput (including random number generation, phase-space mapping, and scattering amplitudes) for the SM Drell-Yan process $u \overline{u}\to e^+ e^- +3g$ (left) and QCD process $gg\to t\overline{t}+2g$ (right) for varying SIMD instruction sets, varying the number of simultaneously scheduled executions of the standalone executable. The throughput is defined as the total number of evaluated amplitudes divided by the total real world runtime. Each of the two CPUs, Intel Gold Xeon 6326, has 16 physical cores and supports hyper-threading for a total of 32 logical units per socket. Regions with hyper-threading (HT) and overcommitment (OC) are marked. For each run $n$ standalone executions are simultaneously scheduled to run $2^{18}$ phase-space points each, and each point is given by the average throughput of 10 such runs. Standard deviations are highlighted albeit typically too small to see.}
    \label{fig:standalone_throughput_icelake_mp}
\end{figure}

As discussed in \cref{sec:technicalevgen}, the multi-event interface now part of \mg{} allows for multithreading alongside SIMD and SIMT parallelism, and to illustrate how these two methods of data-parallelism act orthogonally to each other the \cudacpp{} standalone application also permits multithreading across scattering amplitudes using OpenMP \cite{openmp}. The throughput is expected to increase linearly with both the number of simultaneous running threads and with the SIMD register width, which is tested and shown for two different processes and two different CPUs in \cref{fig:standalone_throughput_skylake,fig:standalone_throughput_icelake}. Here, the throughput is measured as the ratio between the number of evaluated phase-space points and the total amount of time spent on random number generation and scattering amplitude evaluations.

The processes tested here are Drell-Yan plus gluon jets: $u \overline{u} \to e^+ e^- +3g$; and $t\overline{t}$-production: $g g \to t \overline{t} +2g$, both in the LO standard model. The CPUs tested are the Intel Xeon Gold 5118 and the Intel Xeon Gold 6326 . Results for these machines are shown in \cref{fig:standalone_throughput_skylake,fig:standalone_throughput_icelake}, respectively. All tests were run on the Manneback cluster hosted by CISM at UCLouvain on exclusive nodes with two CPUs (in separate sockets) per node.

Throughput generally speeds up following the expected linear growth with number of running threads (up to the number of available threads) and SIMD register width (for implemented instruction sets), although there are some caveats. Across all tests in \cref{fig:standalone_throughput_skylake,fig:standalone_throughput_icelake}, a throughput plateau or even dip is seen when scheduling threads across more threads than are available per socket. A likely culprit here is memory access latency --- although the sockets are equipped with their own RAM, the multithreaded standalone application will land on a specific core and schedule the work across all available threads; it is thus likely that when scheduling across sockets, RAM is still prioritised based on the driving executable itself. While this could be mitigated by runtime parameters, further investigation is low-priority for one simple reason: \textsc{MadEvent} event generation does not utilise executable-end multithreading. Instead, workers are spawned across available threads to run different subprocesses. To test how such multiprocessing affects the throughput, the total throughput when running $n$ individual standalone executables each evaluating $2^{18}$ phase-space points was measured and the results can be seen in \cref{fig:standalone_throughput_skylake_mp,fig:standalone_throughput_icelake_mp}.

\Cref{fig:standalone_throughput_skylake_mp,fig:standalone_throughput_icelake_mp} show that multiprocessing overcomes this cross-socket scheduling issue with no significant implementation overhead --- in fact, the measurements done for these figures are somewhat unfavourable, in that executable launch time is now included in the time measurements --- justifying the choice of multiprocessing for multi-core utilisation in \textsc{MadEvent}. The only caveat to note here is that multiprocessing appears to limit the applicability of hyper-threading. The multithreaded tests in \cref{fig:standalone_throughput_skylake,fig:standalone_throughput_icelake} illustrate a clear speed-up into the hyper-threaded regime (for numbers of threads between one and two times the number of physical cores), which is not achieved with multiprocessing. However, hyper-threading also appears to provide decreasing benefits with SIMD register size, making the benefit marginal. Whether there is any real benefit to gain from multithreading in single-socket machines is left up to future work.

There is an important anomaly in \cref{fig:standalone_throughput_skylake,fig:standalone_throughput_skylake_mp}: AVX-512 does not show any major speed-up when compared to AVX2. This ``non-performant AVX-512'' is a reoccurring phenomenon observed across a range of Intel CPUs while testing \cudacpp{}, and more generally across a wide range of software, pointing towards a hardware issue.

While all current Intel CPUs have two distinct execution ports for 256-bit instructions, so-called ``client'' machines only have one for 512-bit instructions \cite{intel_optimisation_manual}. Consequently, assuming corresponding AVX2 and AVX-512 instructions take the same number of clock cycles, machines like the Gold 5118 will run two AVX2 instructions in the same number of cycles as it runs one AVX-512 instruction. Furthermore, the clock speed is generally decreased when running AVX-512 instructions, meaning the same code on the same machine might end up running slower when compiled with AVX-512 instructions rather than AVX2 --- something that has, in fact, been observed during these tests. To mitigate this, \cudacpp{} has two distinct AVX-512 implementations --- referred to as \texttt{512y} and \texttt{512z} --- where the former actually compiles for 256-bit SIMD registers akin to AVX2 but with additional AVX-512 features, while the latter compiles for 512-bit SIMD registers, just as detailed in \cref{sec:hardware}. All plots here utilise the most performant mode for the given hardware as denoted in the legends of each figure. For details on how to choose which mode to use in \cudacpp{}, refer to \cref{appendix:manual}.

\subsection{SIMT GPU results}
\label{sec:standalone_gpu}

\begin{figure}[t]
    \centering
    \includegraphics[width=0.45\linewidth]{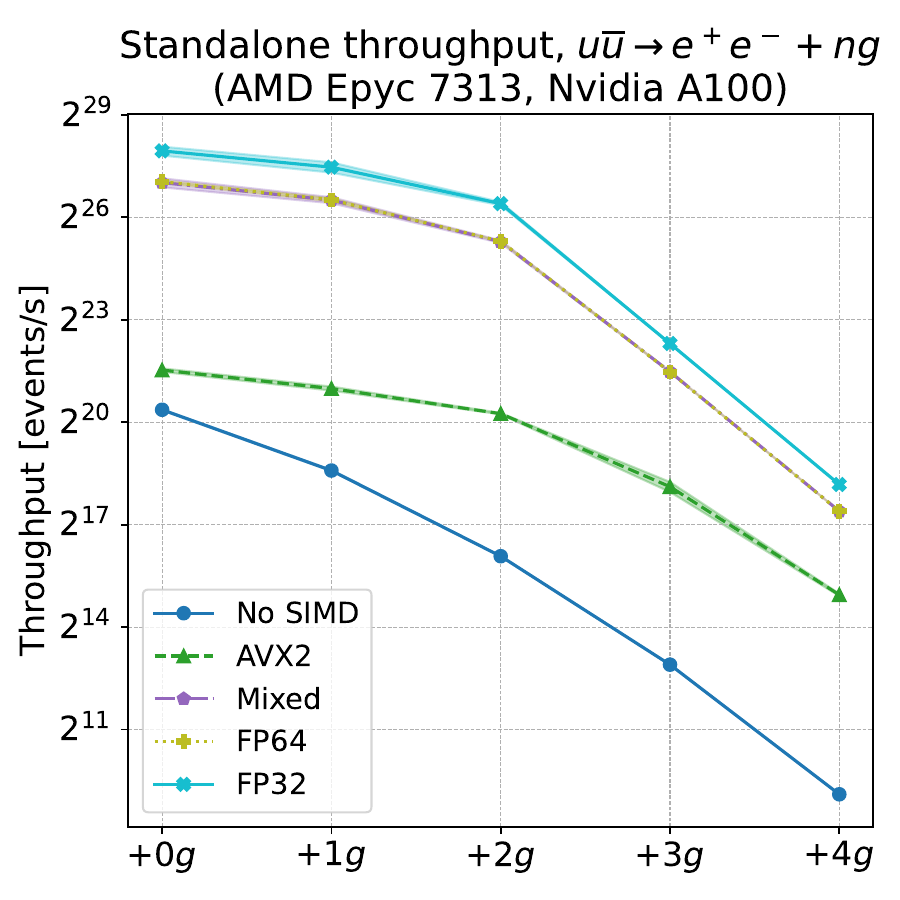}
    \includegraphics[width=0.45\linewidth]{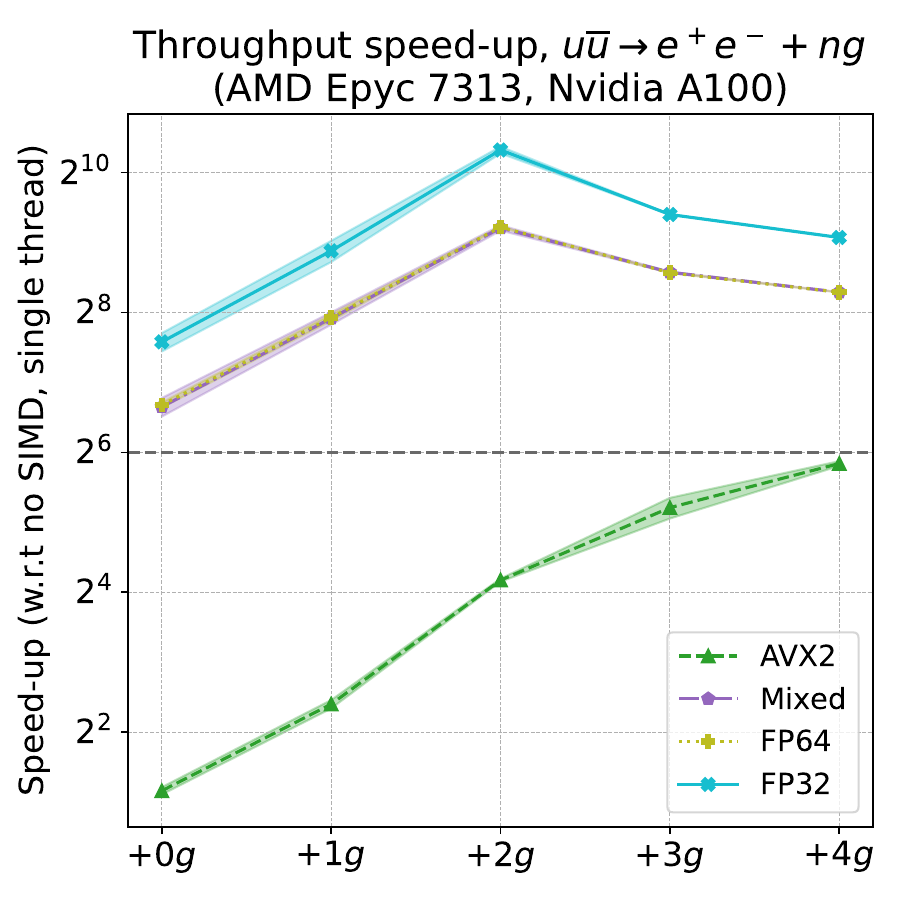}
    \caption{\cudacpp{} standalone throughput (including random number generation, phase-space mapping, and scattering amplitudes) for the SM Drell-Yan process $u\overline{u}\to e^+ e^-$ plus $n\in[0,4]$ gluon jets with differing floating point types running on an Nvidia A100 GPU, as well as comparison with the minimal vectorisation (no SIMD and no multithreading) and maximal vectorisation (AVX2 and multithreading across 16 cores) available on the host CPU (an AMD Epyc 7313) running in mixed precision. The left-hand graph illustrates the wall time throughput, while the right-hand one shows the acceleration as the throughput ratio between the given backend and the unvectorised single-threaded baseline alongside the expected maximal speed-up for the AVX2 line. Each point is given by the average throughput from 10 \cudacpp{} standalone runs on the GPU (CPU) with $2^{26}$ ($2^{18}$) phase-space points each, and standard deviations are highlighted.}
    \label{fig:DY_throughput_gpu}
\end{figure}

\begin{figure}[t]
    \centering
    \includegraphics[width=0.45\linewidth]{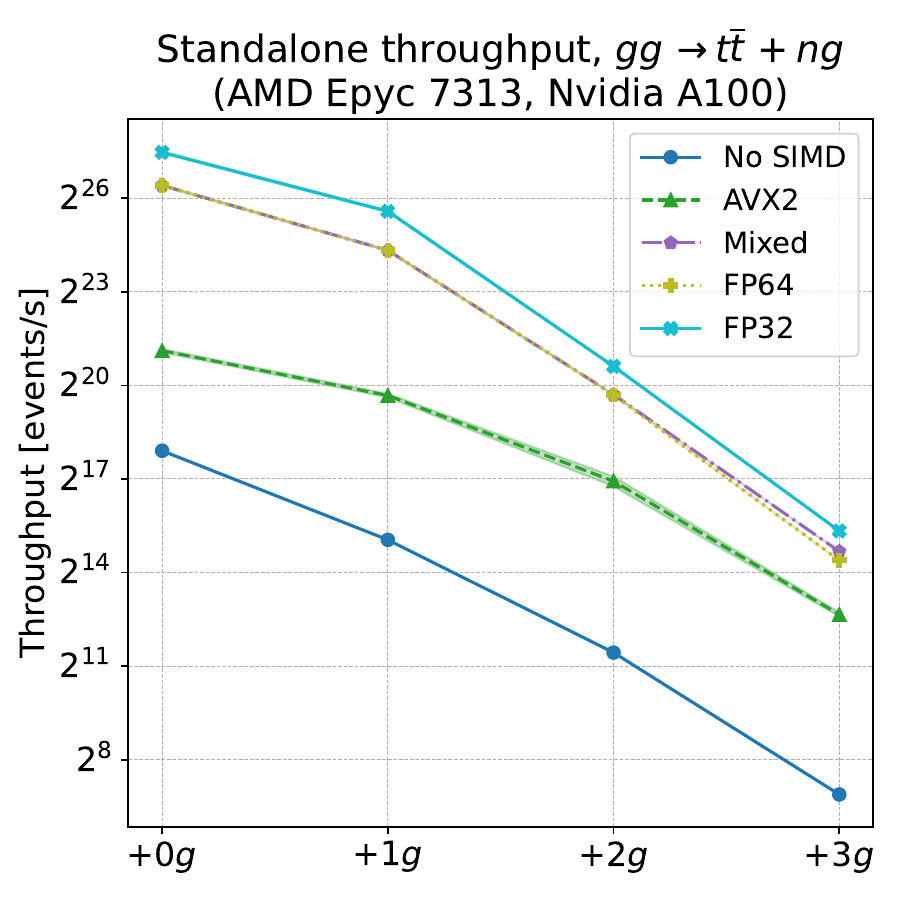}
    \includegraphics[width=0.45\linewidth]{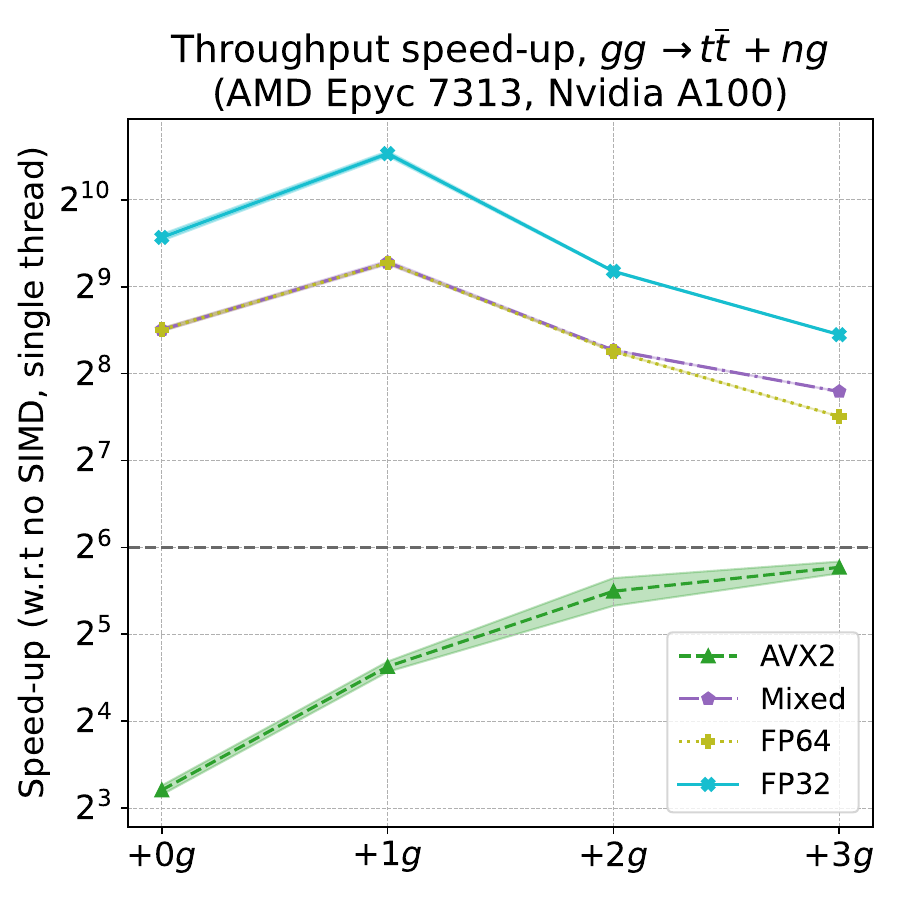}
    \caption{\cudacpp{} standalone throughput (including random number generation, phase-space mapping, and scattering amplitudes) for the SM QCD process $gg\to t\overline{t}$ plus $n\in[0,3]$ gluon jets with differing floating point types running on an Nvidia A100 GPU, as well as comparison with the minimal vectorisation (no SIMD and no multithreading) and maximal vectorisation (AVX2 and multithreading across 16 cores) available on the host CPU (AMD Epyc 7313) running in mixed precision. The left-hand graph illustrates the wall time throughput, while the right-hand one shows the acceleration as the throughput ratio between the given backend and the unvectorised single-threaded baseline. Each point is given by the average throughput from 10 \cudacpp{} standalone runs on the GPU (CPU) with $2^{26}$ ($2^{18}$) phase-space points each, and standard deviations are highlighted.}
    \label{fig:QCD_throughput_gpu}
\end{figure}

Having illustrated that sufficiently complicated \cudacpp{} scattering amplitudes scale optimally with respect to SIMD instruction sets and on-host multithreading in \cref{sec:standalone_simd}, it is time to consider GPU off-loading. As discussed in \cref{sec:hardware,sec:precision} it cannot yet be said whether FP32 would be sufficient for LO helicity amplitudes, limiting \cudacpp{} to higher-end HPC-grade GPUs where the number of floating point operations per second of FP32 and FP64 typically differ by a factor two. Thus, the standalone application should gain up to a factor two speed-up when compiled for FP32 compared to FP64, and the mixed precision mode detailed in \cref{sec:precision} should lie somewhere between these two depending on the complexity of process-specific colour algebra. Note that mixed precision can be used for SIMD parallelism just as well as for SIMT, and is also the default precision mode for \cudacpp{} SIMD compilation, but that the numerical setup is slightly more convoluted --- to make use of the higher throughput for lower precision numbers in a SIMD register, one colour matrix multiplication needs to be performed using two sequential evaluations of kinematic helicity amplitudes which are done in FP64, limiting the immediate acceleration from mixed precision. For SIMT parallelism, though, mixed precision acceleration is immediately accessible as the higher FP32 throughput is a direct consequence of the higher number of FP32 processing units.

Throughput tests for $u \overline{u} \to e^+ e^-$ and $gg \to t\overline{t}$ with increasing numbers of additional gluon jets are shown in \cref{fig:DY_throughput_gpu,fig:QCD_throughput_gpu} respectively, running on an Nvidia A100 GPU with an AMD Epyc 7313 host CPU. The left-hand plots illustrate the throughput for the three compilation modes on the GPU, as well as a ``best'' and ``worst'' case comparison for the host CPU on its own (completely sequential code with no multithreading and AVX2 instructions with multithreading across all 16 physical cores, respectively). The right-hand plots illustrate the ratio between the throughputs for the single-threaded sequential runs and the given backends, i.e. the acceleration relative to the completely sequential code. All illustrated points are mean values from ten separate measurements with the throughput standard deviations highlighted.

The same trends can be seen across \cref{fig:DY_throughput_gpu,fig:QCD_throughput_gpu}. All backends see a significant throughput decrease as the number of external partons increase, in line with the factorial complexity scaling of Feynman diagram methods. It is worth mentioning that the GPU codes and the sequential baseline have very similar trends with process complexity, even if the exact ratios vary, while the multithreaded AVX2 code only starts to approach the expected performance ratio limit of roughly $4\times16 = 2^6$ (a factor 4 from AVX2 and another factor 16 from multithreading) in the high complexity limit, which is due to the throughput definition used here including runtime for random number generation and mapping (which do not benefit from SIMD instructions) and from the inter-thread latency limiting any speed-up for very simple tasks. Additionally, it deserves mention that consistently achieving maximal throughput on the A100 requires a factor $2^8=256$ times more events than for the SIMD CPU case.

For almost all processes considered in \cref{fig:DY_throughput_gpu,fig:QCD_throughput_gpu} the difference between FP64 and mixed precision throughput is negligible. This is not surprising; while the colour algebra complexity scales as $\mathcal{O}\left( (n!)^2\right)$ with $n$ the number of strongly interacting external particles, the calculation itself is a single matrix multiplication which only becomes substantial as the size of the colour matrix grows significant. However, in the rightmost point in \cref{fig:QCD_throughput_gpu} representing the process $gg\to t \overline{t} +3g$ the difference becomes sizeable: in FP32 and FP64, the colour algebra makes up roughly a third of scattering amplitude compute time, and mixed precision calculations end up with $\sim22\%$ greater throughput than FP64.

\section{Leading-order event generation}
\label{sec:evgen}

While \cref{sec:standalone} exposes the full data-parallelism speed-up in helicity amplitudes, which for SIMD CPUs scales linearly with SIMD register size, this does not directly translate to event generation. The impact of data-parallelism on runtime is, of course, dependent on the runtime fraction made up of the code parallelised. In \cref{sec:madevent_runtime} the compute makeup of LO event generation in \textsc{MadEvent} is demonstrated across a variety of SM processes with varying complexity, and the possible acceleration provided by data-parallel helicity amplitudes is discussed in the context of Amdahl's law. Then, in \cref{sec:evgen_simd,sec:evgen_gpu} the real acceleration is investigated for SIMD CPUs and SIMT GPUs, respectively. Before going into these topics, though, it is best to give an overview of the way these tests were performed, so as to not have to reiterate the details in both \cref{sec:evgen_simd,sec:evgen_gpu}.

To start, two types of processes are considered: single-channel processes, where one specific configuration of external partons is considered, and multi-channel processes, where the initial state can be composed of any proton partons and final state jets can be any massless QCD particle. Single-channel processes are the type considered for the standalone application in \cref{sec:standalone}, whereas multi-channel processes are the linear combinations that would be observed in a real-world collider. All processes considered are inclusive but at fixed parton-level multiplicities (i.e. the total number of external partons is fixed within \mg{}), but merged process runtimes should be linear combinations of ones with differing multiplicities. Again, just as in \cref{sec:standalone}, the processes considered are Drell-Yan plus jets and pure QCD $t\overline{t}$-production in the SM at LHC energies.

Runtimes are measured using gridpacks, minimal pre-compiled \textsc{MadEvent} instances with preset \textit{grids}, i.e. phase-space splittings according to different regions' relevance to observable measurements; this essentially means that unweighting efficiency will be optimal, minimising the number of phase-space points evaluated per unweighted event. Additionally, runtimes are measured in terms of the time it takes to generate some number of unweighted events, making the total number of phase-space points evaluated stochastic and including unweighting and LHE file output into the total runtimes. Unless otherwise specified, the optional gridpack runtime flags \texttt{-p} and \texttt{-m} described in \cref{sec:gridpacks} will take the default values of 1 and 2 500, respectively.

Finally, all tests are run without helicity recycling \cite{Mattelaer:2021xdr} and without the default \mg{} parton grouping, as neither are supported in \cudacpp{} as of yet. The former involves major runtime modifications to amplitude routines, while the latter uses branching structures to combine partons which have similar but non-identical kinematic structures. Parton grouping has little impact on runtime, but helicity recycling can speed up computations by more than a factor 2 for processes with many external vector bosons. These effects are completely omitted from Fortran comparisons below. Note that these both have large impacts on compile time.

\subsection{Event generation runtime makeup}
\label{sec:madevent_runtime}

A plethora of profiling tools have been used both for \mg{} in general and \cudacpp{} development in particular, including but not limited to Valgrind \cite{10.1145/1273442.1250746}, Callgrind \cite{10.1007/978-3-540-24688-6_58}, perf \cite{perfwikiPerfLinux}, Adaptyst \cite{graczyk2025enhancingsoftwarehardwarecodesignhep}, and various internally developed scripts and software. Of particular use in determining compute bottlenecks are FlameGraphs \cite{brendangreggFlameGraphs}, a tool for unwinding and graphically representing callgraphs in an easily parsed visual format. Starting from the profiled execution, all calls to the same function are collapsed into a single block such that the magnitude along the $x$-axis represents fraction of events which were recorded in the given function and the $y$-axis reflects stack depth; e.g., if 50\% of runtime is spent in the function \texttt{bar} called by the function \texttt{foo} called by the profiled program \texttt{main}, the \texttt{bar} block would lie on top of the \texttt{foo} block on top of the \texttt{main} block, and \texttt{bar} would stretch across half of the FlameGraph. Function calls from a given block are ordered alphabetically.

\begin{figure}[t]
    \centering
    \includegraphics[width=0.95\linewidth]{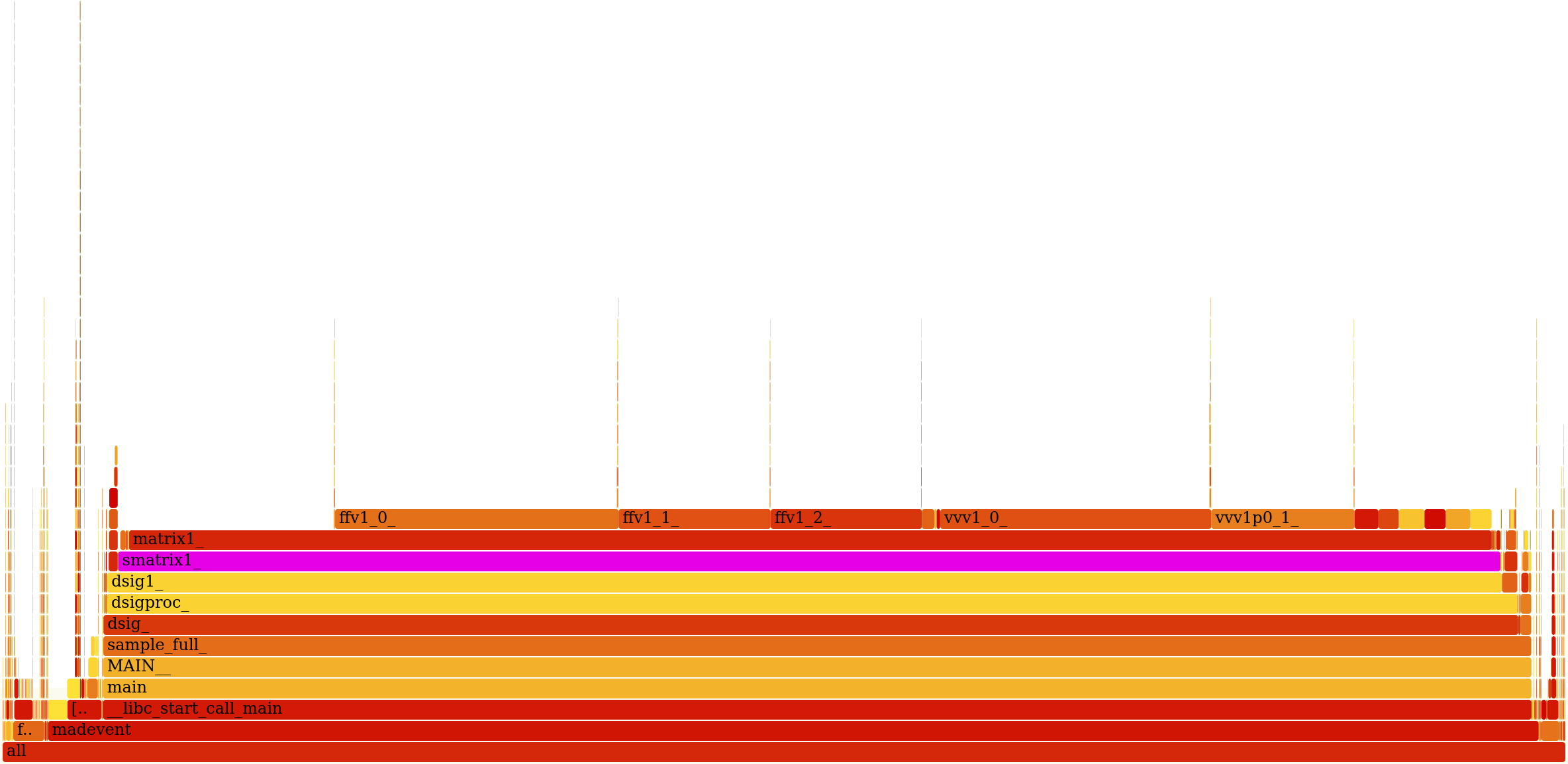}
    \caption{FlameGraph \cite{brendangreggFlameGraphs} of Fortran \textsc{MadEvent} event generation for the process $gg\to t \overline{t} \, gg$ without helicity recycling \cite{Mattelaer:2021xdr}, generating $10\, 000$ unweighted events. The subroutine \texttt{smatrix1}, which wraps helicity amplitude evaluations, has been highlighted. The $x$-axis displays runtime fraction (called routines are listed in alphabetical order) while the $y$-axis corresponds to stack depth. The lowest stack, \texttt{all}, reflects the entirety of profiled execution; the second lowest stack includes \texttt{f951} (barely visible), the symbol for the \texttt{gfortran} compiler, as well as \texttt{madevent}, the Python \textsc{MadEvent} driver. The ``actual'' Fortran executable is given by \texttt{main}, which is what runtime percentages below are calculated with respect to.}
    \label{fig:flamegraph}
\end{figure}

\Cref{fig:flamegraph} displays a FlameGraph of \textsc{MadEvent} leading order event generation for the process $gg\to t \overline{t} \, gg$ without helicity recycling, representing the type of process-specific profiling used below. In this particular figure, the Python \texttt{madevent} driver makes $95.4\%$ of total runtime, but e.g. \texttt{f951} (the symbol for \texttt{gfortran}) is barely visible to its right, making up $1.98\%$ of runtime here. The purple, highlighted block \texttt{smatrix1} is the entry point for helicity amplitude evaluations, i.e. the part of the program that has been parallelised in \cudacpp{}, and it makes up $88.4\%$ of total runtime for this process. However, the more relevant comparison to make is between \texttt{smatrix1} and either \texttt{madevent} or \texttt{main}, which are the Python driver and the Fortran executable, respectively. Here, the latter is used, giving \texttt{smatrix1} a relative runtime fraction of $96.7\%$

\begin{table}[t]
    \centering
    \begin{tabular}{|c|c|c|c|} \hline
        No. jets &  $e^+ e^- \to \mu^+ \mu^- + n\gamma$ & $u \overline{u} \to e^+ e^- + ng$ &$gg\to t\overline{t} + ng$  \\ \hline
        $n=0$ & $20.9\%$ & \phantom{0}$9.3\%$ & $34.6\%$ \\ \hline
        $n=1$ & $43.6\%$ & $12.7\%$ & $81.3\%$  \\ \hline
        $n=2$ & $63.7\%$ & $21.1\%$ & $96.7\%$  \\ \hline
        $n=3$ & $86.6\%$ & $56.7\%$ & $99.6\%$ \\ \hline
    \end{tabular}
    \caption{Fraction of \textsc{MadEvent} Fortran executable spent in helicity amplitudes for processes $e^+e^-\to\mu^+\mu^-$, $u\overline{u}\to e^+e^-$, and $gg \to t\overline{t}$, plus $n$ additional final state photons, gluons, and gluons respectively, when generating 10 000 unweighted events. While only fraction of individual runtimes are shown above, a higher fraction of runtime spent in amplitudes coincides with an increased total runtime; for example, generating $10\,000$ unweighted events for $ u\overline{u} \to e^+ e^- + 3g$ took roughly 10 CPU minutes, whereas generating $10\,000$ unweighted events for $gg\to t\overline{t} + 3g$ took roughly 40 CPU hours.}
    \label{tab:flamegraph_summary}
\end{table}

A summary of the runtime fraction spent in helicity amplitudes for various different processes are provided in \cref{tab:flamegraph_summary}, which shows the exact expected trend. As process complexity increases, the runtime fraction spent in the \texttt{smatrix1} routine does as well.\footnote{Note that the central column, displaying the Drell-Yan process $u \overline{u} \to e^+ e^- + ng$, scales a lot slower with additional jets (here, gluons) than the other two. This follows from the fact that this process mixes EW and QCD interactions; the factorial scaling of Feynman diagram methods is roughly independent in interaction types.} This is especially apparent for the purely QCD $t \overline{t}$-production process in the right-most column.

\begin{figure}[t]
    \centering
    \includegraphics[width=0.99\linewidth]{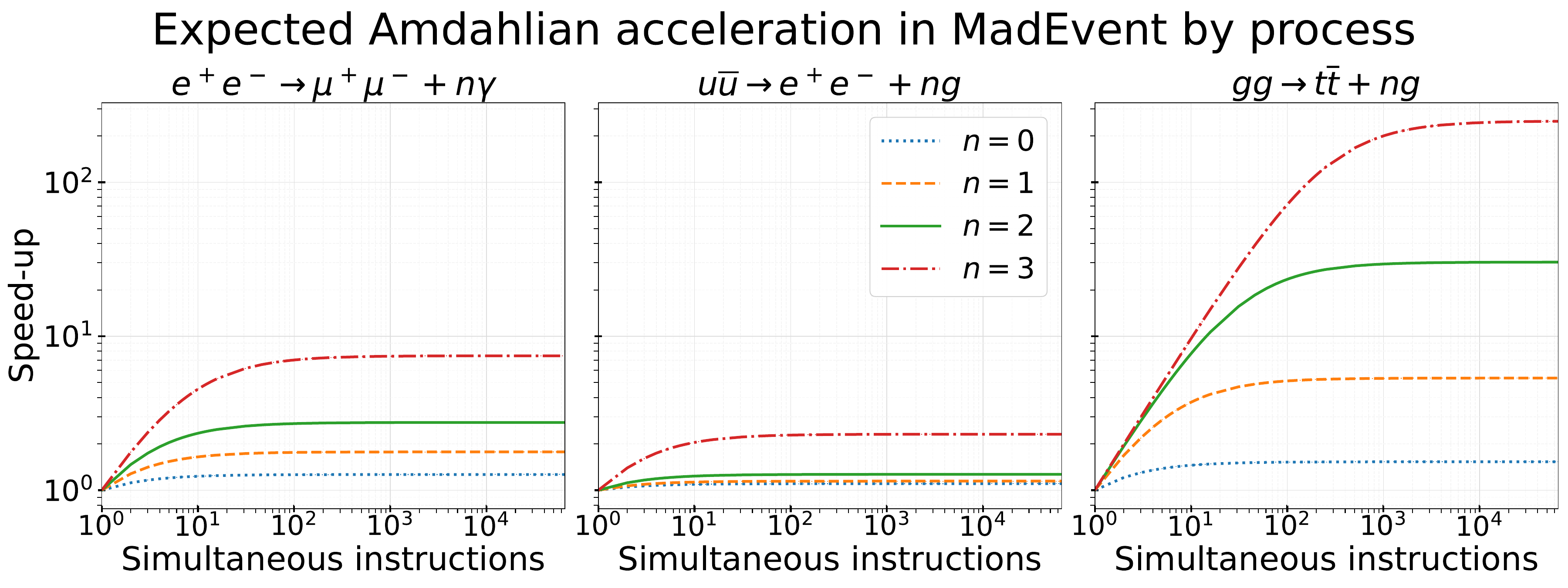}
    \caption{Expected Amdahlian speed-up in \textsc{MadEvent} event generation from data-parallel helicity amplitudes as a function of the number of simultaneous instructions for processes $e^+e^-\to\mu^+\mu^-$, $u\overline{u}\to e^+e^-$, and $gg \to t\overline{t}$, plus $n$ additional final state photons, gluons, and gluons respectively. Fraction of runtime spent in the parallelisable part of the program (i.e. helicity amplitudes) taken from \cref{tab:flamegraph_summary}.}
    \label{fig:amdahl_mg}
\end{figure}

Given a program that processes some amount $n$ data in $t$ time, runtime acceleration can be defined as
\begin{align} \label{eq:acceleration}
    \text{Acceleration} = \left( \frac{n_{\text{par.}}}{t_{\text{par.}}} \right)\left/ \left( \frac{n_{\text{seq.}}}{t_{\text{seq.}}} \right)   \right. =  \left( \frac{n_{\text{par.}}}{n_{\text{seq.}}} \right) \times  \left( \frac{t_{\text{seq.}}}{t_{\text{par.}}} \right),
\end{align}
where indices refer to parallelised (par.) execution and sequential (seq.) execution. However, this generic formula is not necessarily representative. A fairer comparison is to fix one of the right-hand fractions in \cref{eq:acceleration} to 1, that is, to say the problem either has a fixed size or a fixed runtime. The former, how much faster a given problem size is handled, will here be referred to as the Amdahlian acceleration; the latter, how much more data fits into a given runtime, will be called the Gustafsonian acceleration. Below, the Amdahlian acceleration is considered, but it is worth noting that for a problem with linear complexity scaling (e.g. event generation) the two definitions are equivalent in the large $n$, large $t$ limits.

Amdahlian acceleration can more simply be calculated in terms of the fraction of program runtime $p$ spent in the code section to be parallelised and the number of parallel instructions $n$ as
\begin{align} \label{eq:amdahl}
    \text{Acceleration} \leq \frac{1}{1 - p\left( 1 - \frac{1}{n}  \right)} < \frac{1}{1-p},
\end{align}
i.e. the upper bound of Amdahlian acceleration is given by the fraction of program runtime spent in the remaining \textit{sequential} part of the program. This formula, known as \textit{Amdahl's law}, can be used to forecast \cudacpp{} speed-up, and in \cref{fig:amdahl_mg} graphs depicting projected Amdahlian acceleration for the processes shown in \cref{tab:flamegraph_summary} are given. Just as expected, the projected acceleration increases with process complexity. In particular high-multiplicity QCD processes could be sped up greatly.

\subsection{SIMD CPU results}
\label{sec:evgen_simd}

\begin{figure}[t]
    \centering
    \begin{subfigure}{\textwidth}
    \includegraphics[width=0.99\linewidth]{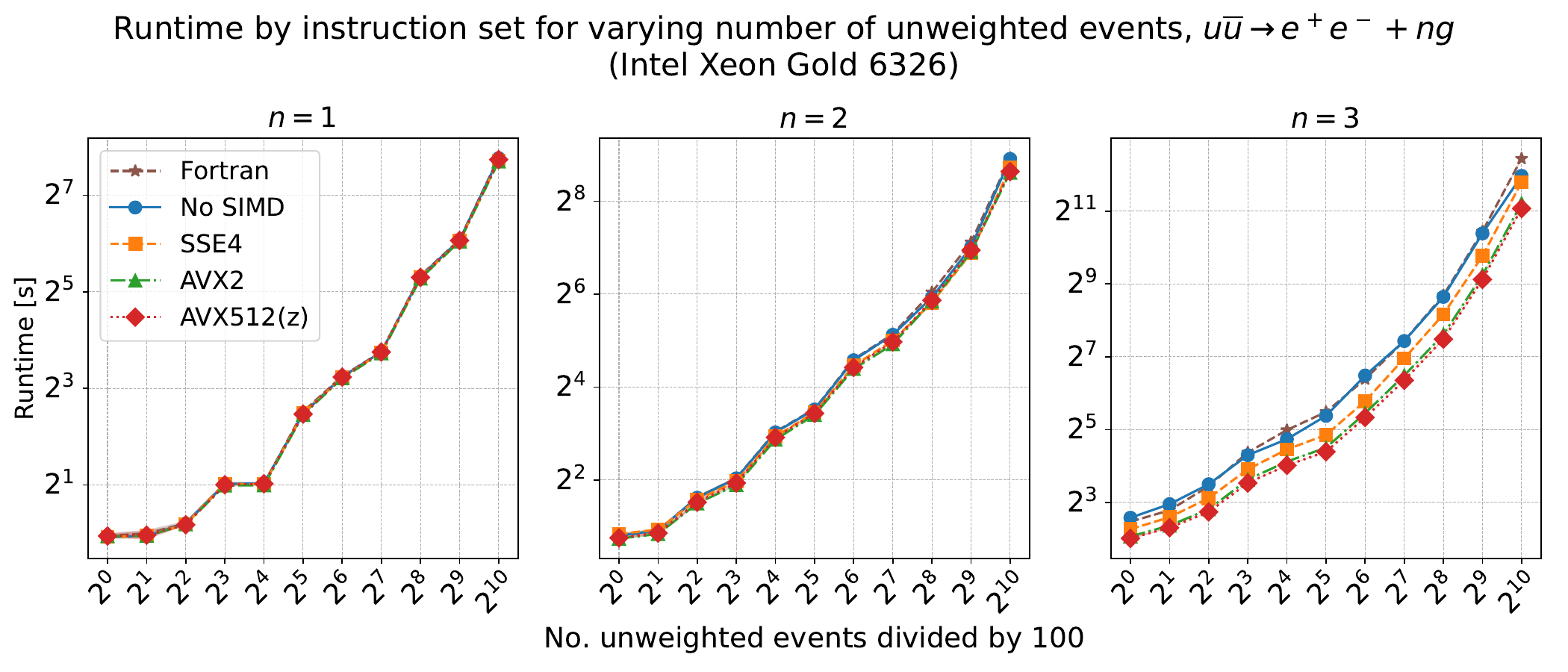}
    \end{subfigure}
    \begin{subfigure}{\textwidth}
        \includegraphics[width=0.99\linewidth]{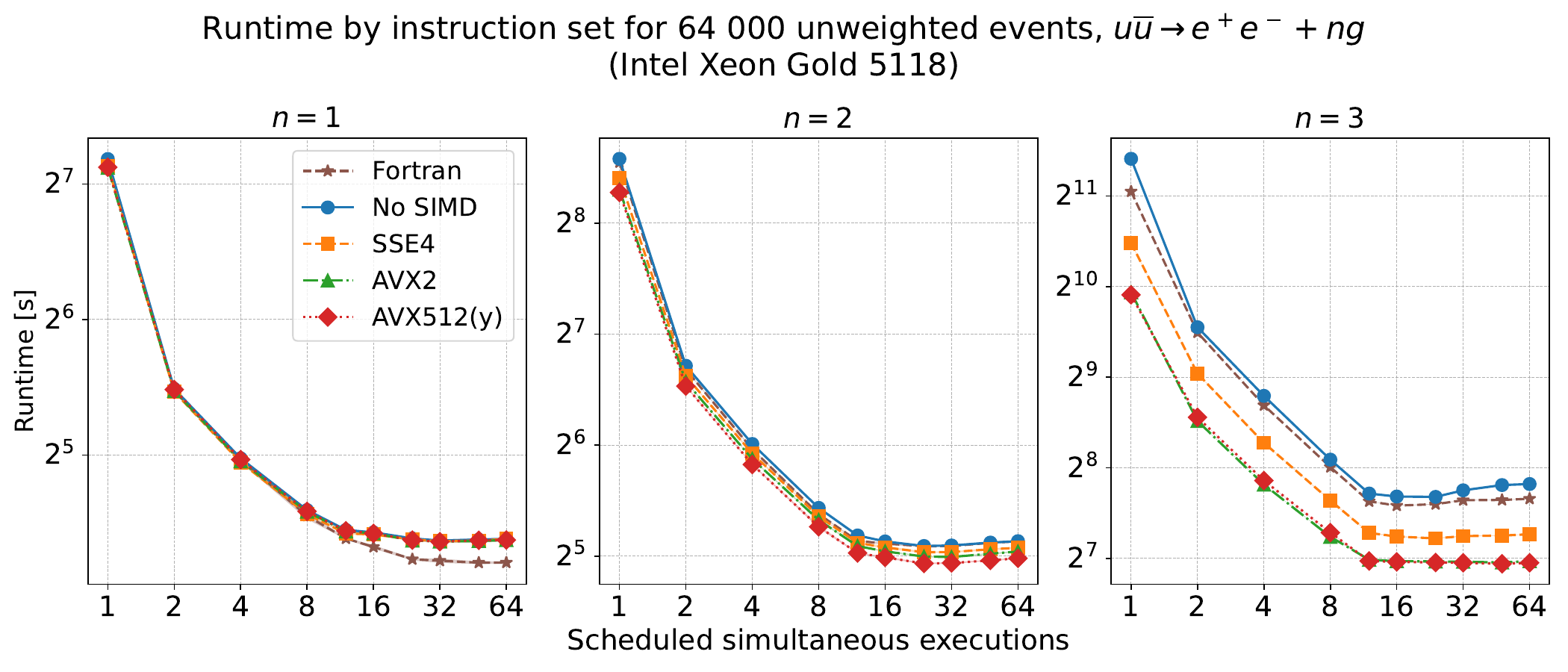}
    \end{subfigure}
    \caption{Total runtime to generate unweighted events for the process $u\overline{u}\to e^+ e^- + ng$ for $1 \leq n \leq3$ at LO in \textsc{MadEvent} with (top) fixed single core execution and varying number of unweighted events, and (bottom) varying number of simultaneous executions scheduled by the Python driver using the gridpack runtime flag \texttt{-p} with fixed number of unweighted events. Points are given by the mean runtime of five subsequent runs and standard deviations are highlighted albeit generally too small to be visible.}
    \label{fig:DY_single_channel}
\end{figure}

\begin{figure}[t]
    \centering
    \begin{subfigure}{\textwidth}
    \includegraphics[width=0.99\linewidth]{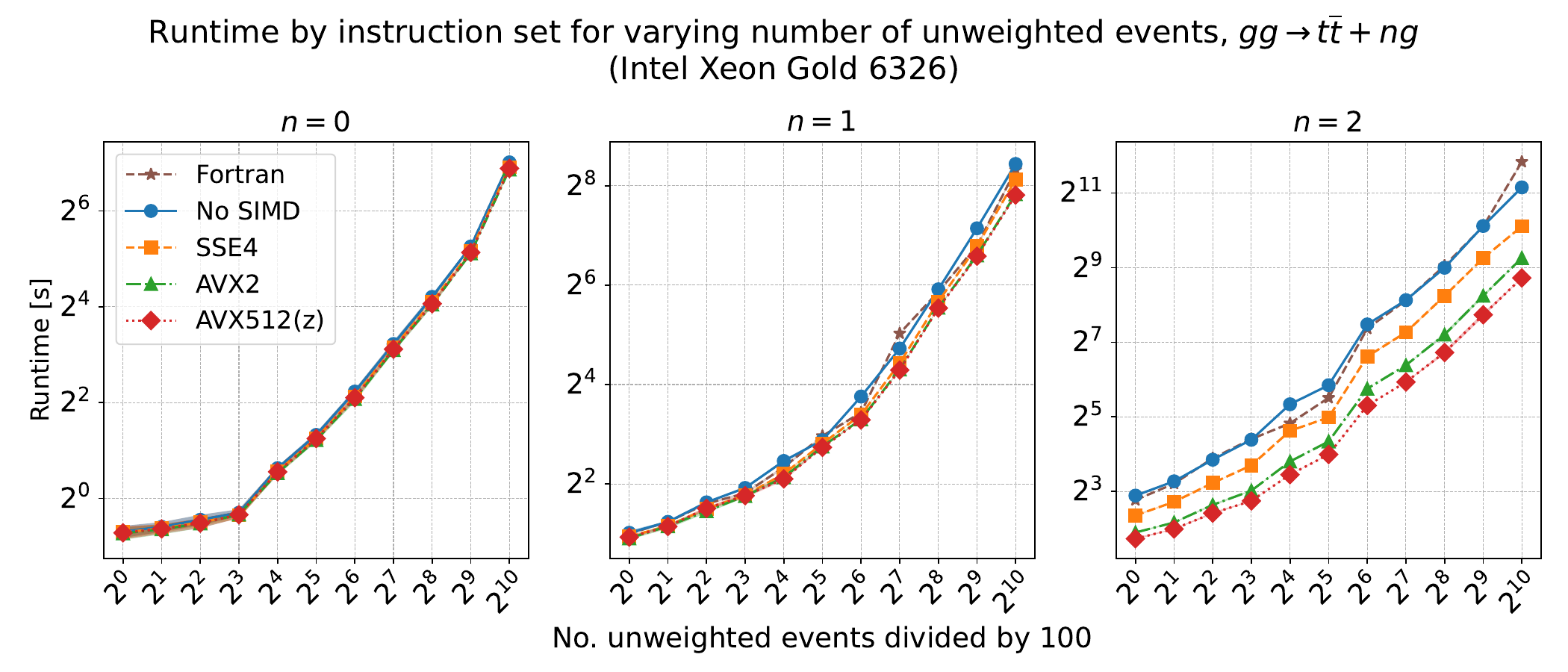}
    \end{subfigure}
    \begin{subfigure}{\textwidth}
        \includegraphics[width=0.99\linewidth]{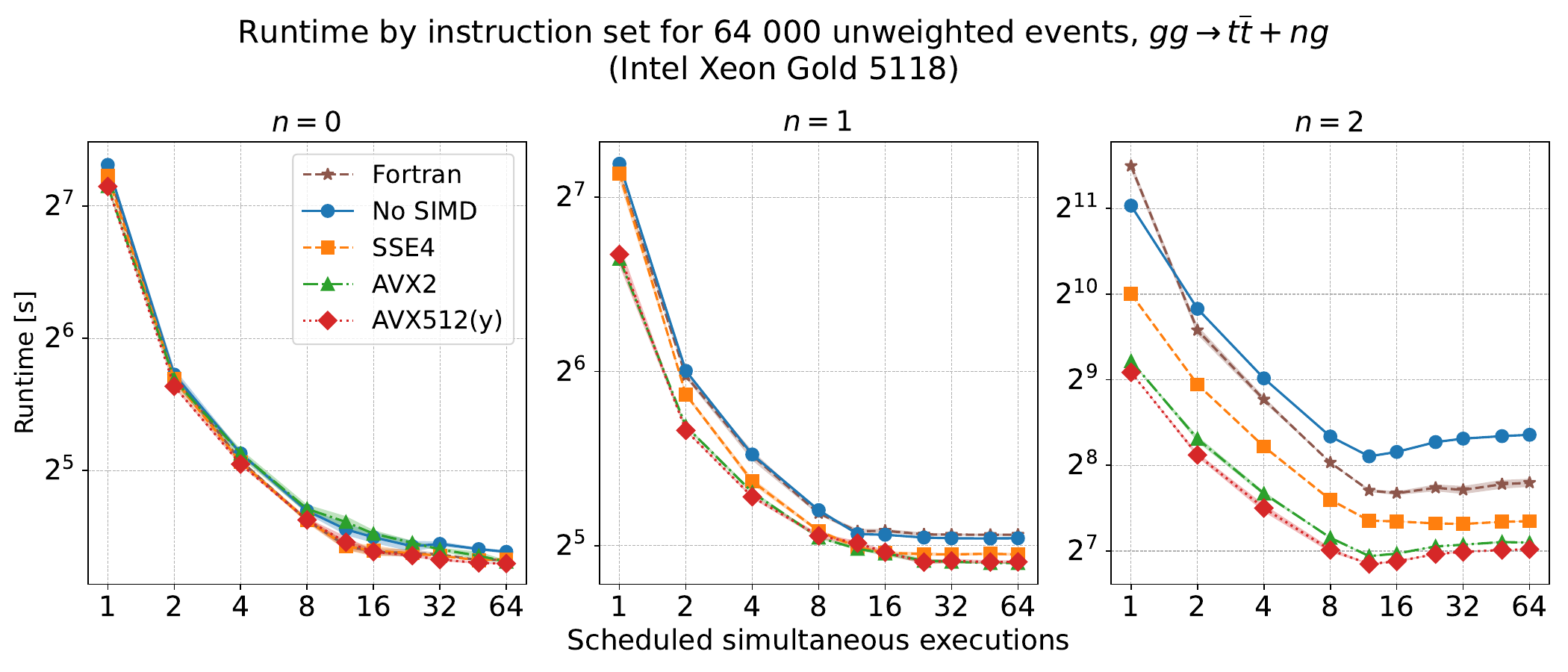}
    \end{subfigure}
    \caption{Total runtime to generate unweighted events for the process $gg\to t \overline{t} + ng$ for $0 \leq n \leq2$ at LO in \textsc{MadEvent} with (top) fixed single core execution and varying number of unweighted events, and (bottom) varying number of simultaneous executions scheduled by the Python driver using the gridpack runtime flag \texttt{-p} with fixed number of unweighted events. Points are given by the mean runtime of five subsequent runs and standard deviations are highlighted albeit generally too small to be visible.}
    \label{fig:QCD_single_channel}
\end{figure}

\begin{figure}[t]
    \centering
    \begin{subfigure}{\textwidth}
    \includegraphics[width=0.99\linewidth]{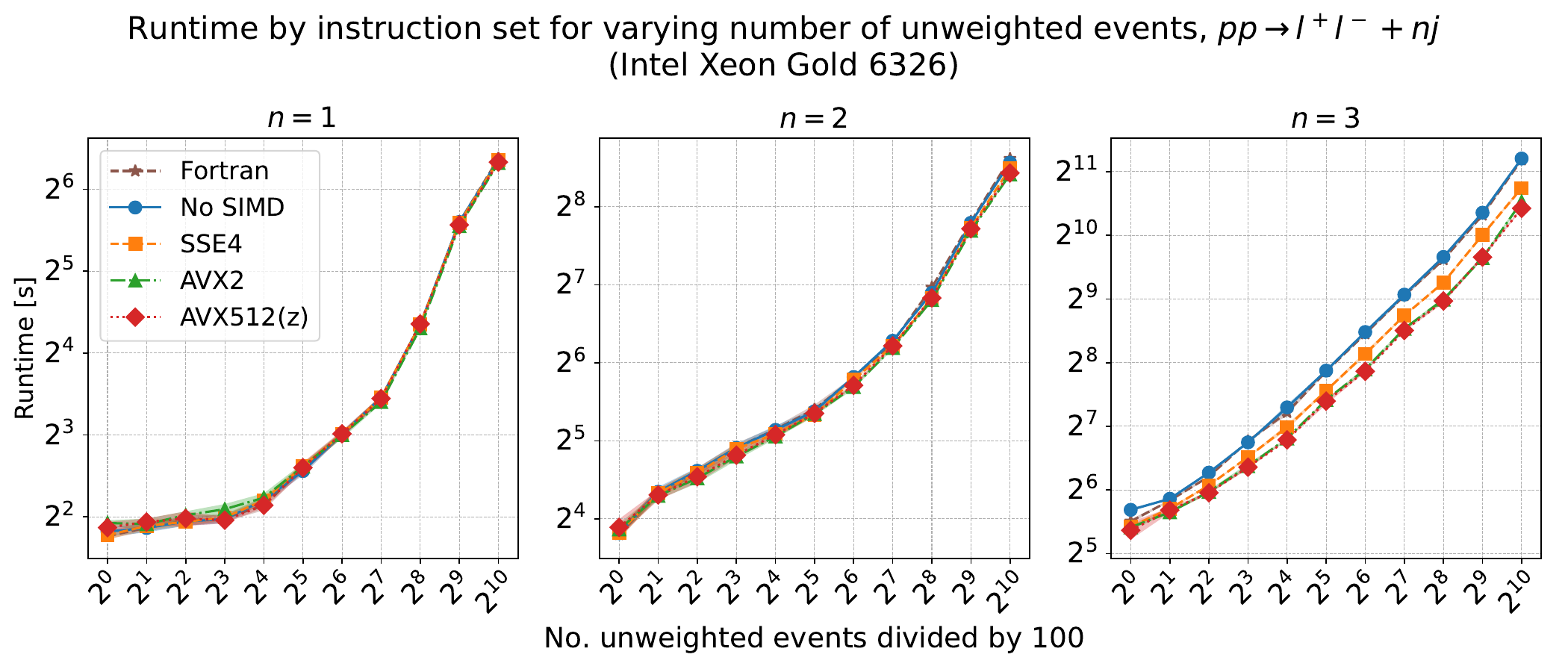}
    \end{subfigure}
    \begin{subfigure}{\textwidth}
        \includegraphics[width=0.99\linewidth]{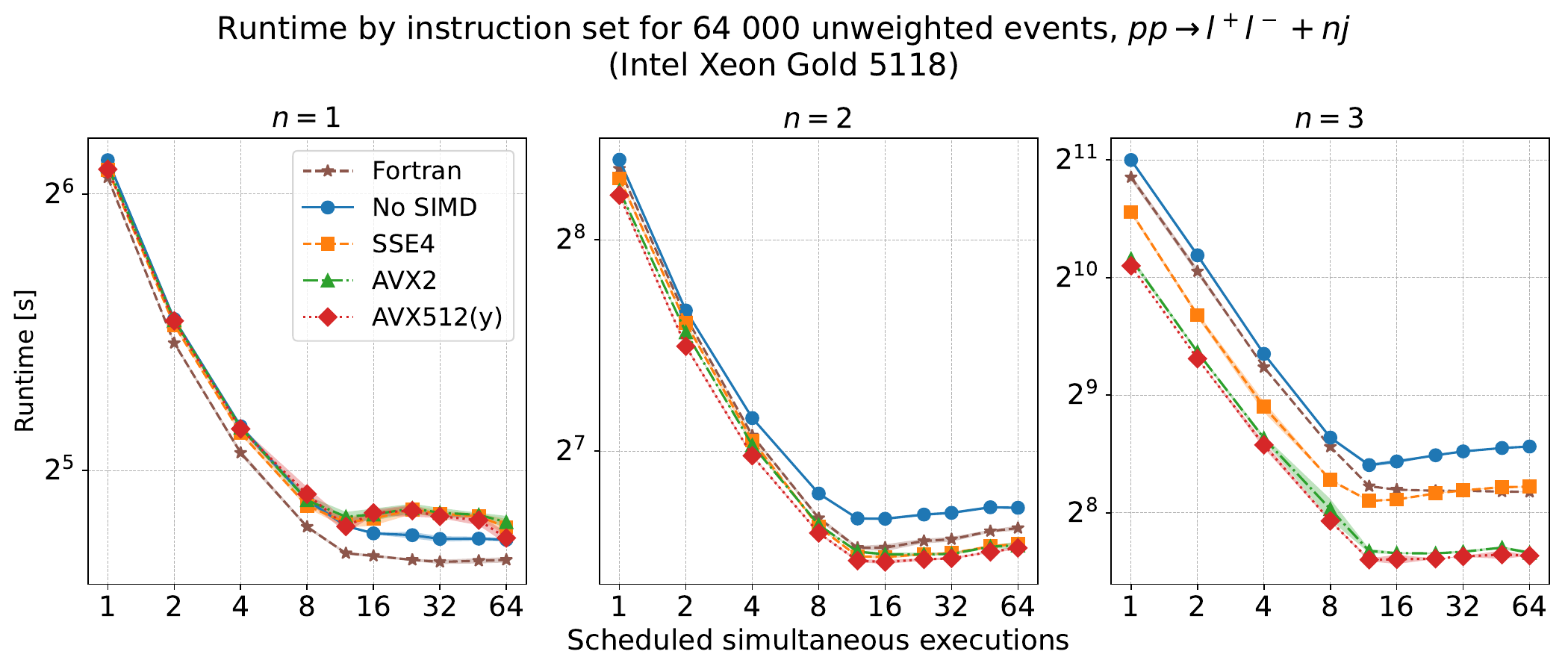}
    \end{subfigure}
    \caption{Total runtime to generate unweighted events for the process $pp\to l^+ l^- + nj$ for $1 \leq n \leq3$ with $j$ any massless QCD jet at LO in \textsc{MadEvent} with (top) fixed single core execution and varying number of unweighted events, and (bottom) varying number of simultaneous executions scheduled by the Python driver using the gridpack runtime flag \texttt{-p} with fixed number of unweighted events. Points are given by the mean runtime of five subsequent runs and standard deviations are highlighted albeit generally too small to be visible.}
    \label{fig:DY_multi_channel}
\end{figure}

\begin{figure}[t]
    \centering
    \begin{subfigure}{\textwidth}
    \includegraphics[width=0.99\linewidth]{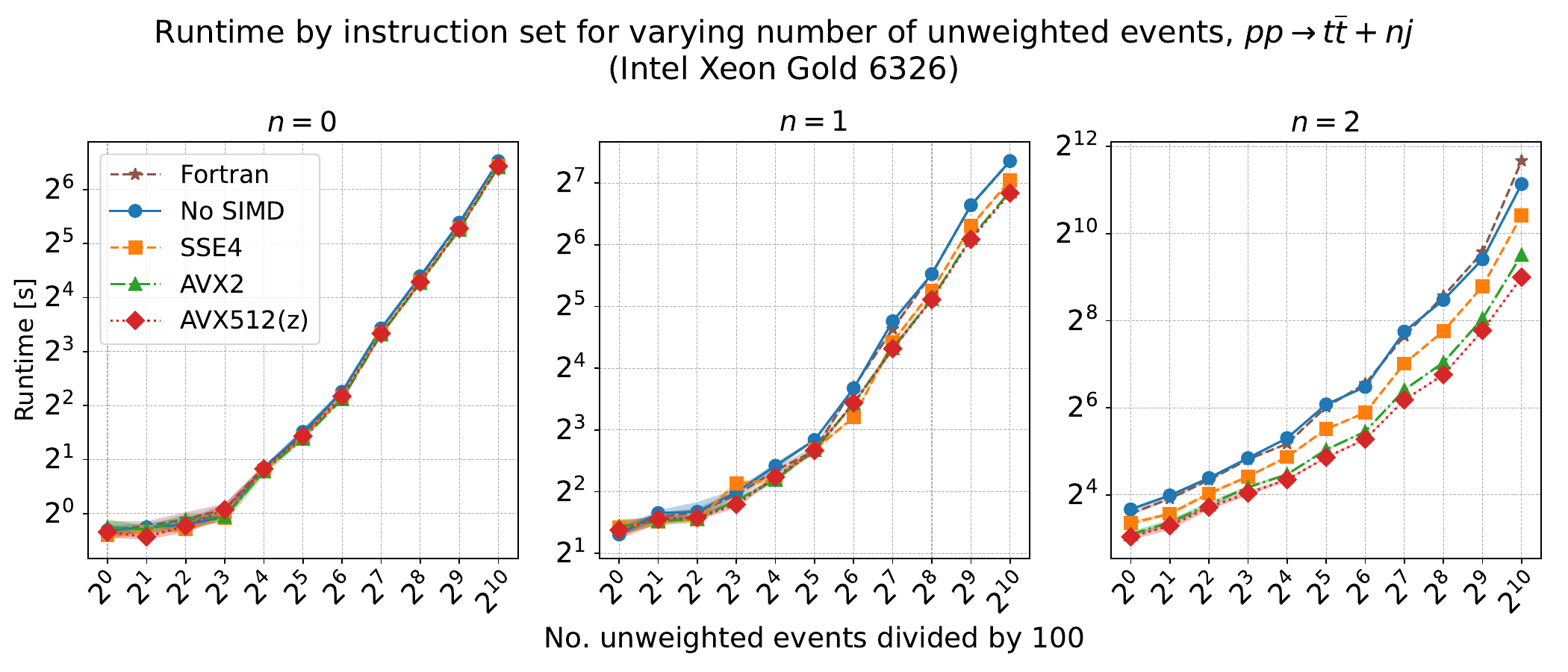}
    \end{subfigure}
    \begin{subfigure}{\textwidth}
        \includegraphics[width=0.99\linewidth]{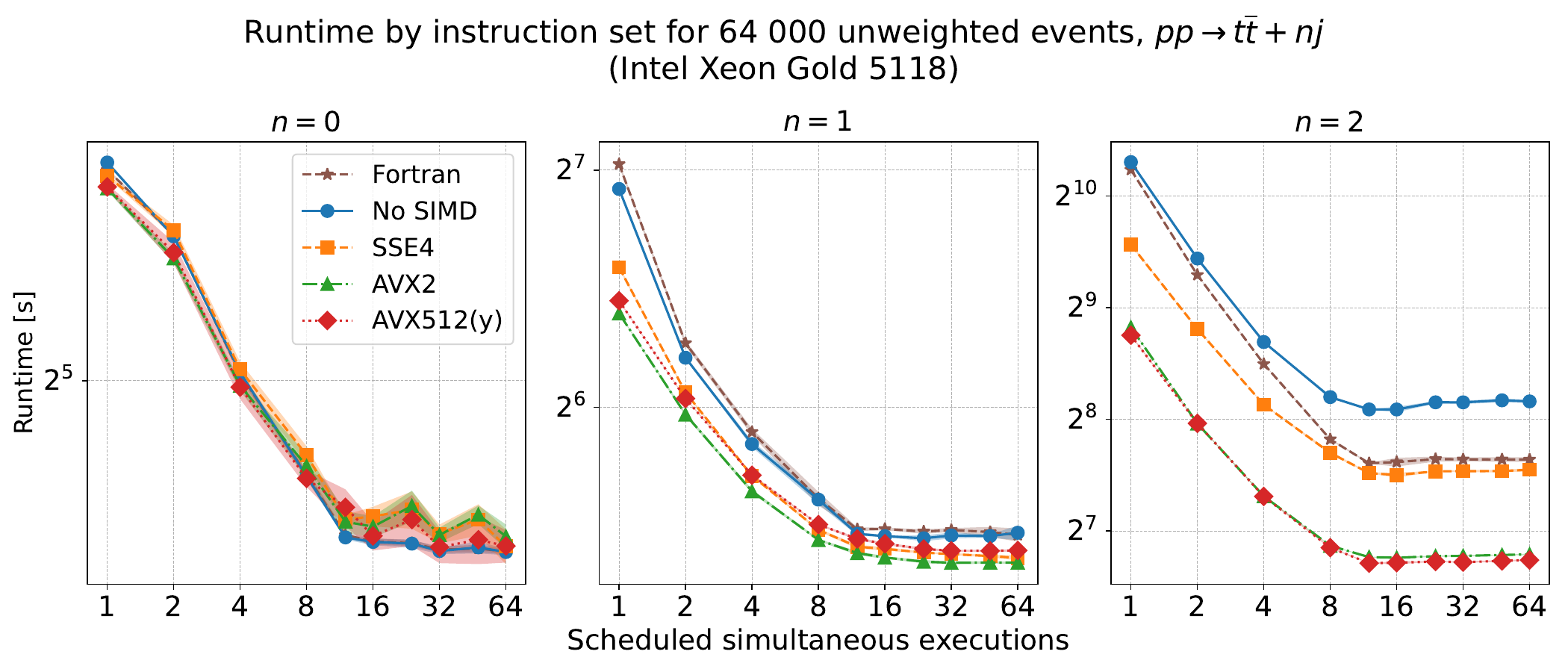}
    \end{subfigure}
    \caption{Total runtime to generate unweighted events for the process $pp\to t\overline{t} + nj$ for $0 \leq n \leq2$ with $j$ any massless QCD jet at LO in \textsc{MadEvent} with (top) fixed single core execution and varying number of unweighted events, and (bottom) varying number of simultaneous executions scheduled by the Python driver using the gridpack runtime flag \texttt{-p} with fixed number of unweighted events. Points are given by the mean runtime of five subsequent runs and standard deviations are highlighted.}
    \label{fig:QCD_multi_channel}
\end{figure}

In consideration of the discussion regarding speed-up from data parallelism in \cref{sec:madevent_runtime}, a better benchmark than optimal throughput is a direct runtime comparison. In \mg{} this can be done with gridpacks --- minimal pre-compiled \textsc{MadEvent} executables for a given process with defined phase-space grids to sample processes. Up until recently \mg{} gridpacks were restricted to running single core, but in order to optimise GPU usage for \cudacpp{} a runtime argument \texttt{--parallel} has been added as described in \cref{sec:gridpacks}, defining the number of simultaneously scheduled \textsc{MadEvent} Fortran executions. Two axes for scaling can now be tested: real-world runtime increase with number of requested unweighted events, and real world runtime decrease with number of scheduled \textsc{MadEvent} executions. All SIMD tests presented in this section were run on the Manneback cluster hosted by CISM at UCLouvain, with one core used for tests on Intel Xeon Gold 6326 CPUs and twelve cores used for tests on Intel Xeon Gold 5118 CPUs.

First, comparisons are made for single-channel processes akin to the standalone tests in \cref{sec:standalone}. In \cref{fig:DY_single_channel,fig:QCD_single_channel} runtimes are shown for the SM processes $u \overline{u}\to e^+ e^- + (n+1)g$ and $g g \to t \overline{t} +ng$, respectively, for $0\leq n\leq2$. The top plots illustrate runtime increase with the number of requested unweighted events across the different SIMD instruction sets on hardware with fully performant AVX-512, while the bottom plots illustrate the runtime decrease with the scheduled number of \textsc{MadEvent} executions with the number of requested events fixed at 64 000 while limiting the total number of available CPU cores to 12 on hardware with non-performant AVX-512 (see \cref{sec:standalone_simd}). Note that the lines denoted Fortran lack helicity recycling and use the \cudacpp{} parton grouping, which differs from the default in \mg{} due to the branching nature of the code for the latter.

The same general trends can be seen across \cref{fig:DY_single_channel,fig:QCD_single_channel}: the top plots illustrate that the impact of data-parallel helicity amplitudes only become apparent for high-complexity processes, and that the full benefit can only be exploited in the many events limit --- aligning with expectations. Whereas the standalone results depicted in \cref{sec:standalone} showed a one-to-one throughput increase with both SIMD register size and utilised thread count, the benefit in event generation is limited by the remaining sequential code sections. The top plots illustrate that the performance increase between no SIMD, SSE4, AVX2 and AVX-512 are no longer exact factors of two, whereas the lower plots show that there is a limit to the gain from multiprocessing (for a fixed number of events with a fixed maximal number of unweighted events generated per executable) as is especially apparent in the low-complexity left-hand plots, although the limit of 12 CPU cores appears insufficient to quite reach this supposed plateau.

However, there are some unexpected behaviours. For example, the sudden runtime growth for Fortran between the final few points in the right-most top plots of both \cref{fig:DY_single_channel,fig:QCD_single_channel} indicate that there is a bottleneck in the many-events limit that is overcome in the \cudacpp{} code regardless of SIMD utilisation. The source of this bottleneck is not immediately clear, and while  these results are more favourable for \cudacpp{} than expected, they are not sufficiently extreme to necessitate immediate investigation; more thorough tests will be performed in the future, but are not high priority.

Another intriguing observation comes from the bottom plots. Just as multithreading provided a linear speed-up with the number of threads for standalone, multiprocessing generally provides a linear speed-up with the number of simultaneous executions, with one major caveat: going from one to two simultaneous executions more than halves the runtime for single-channel processes, consistently. This occurs regardless of backend used, suggesting it is an effect of the multi-core gridpack runs rather than any explicit \cudacpp{} development. However, as shall soon be seen, this benefit does not translate to multi-channel processes, making the behaviour a low-priority curiosity whose investigation will be left for future work.

\Cref{fig:DY_multi_channel,fig:QCD_multi_channel} exhibit the same information as \cref{fig:DY_single_channel,fig:QCD_single_channel} for the corresponding multi-channel processes $p p \to l^+ l^- + (n+1)j$ and $p p \to t \overline{t} + nj$, respectively, where $j$ can be any massless QCD jet and $0 \leq n \leq2$. The same general trends can be seen as in the single-channel cases: the benefits of SIMD instructions only become apparent in the high-complexity many-events limits and the maximal acceleration is limited by remaining sequential code sections; and multiprocessing provides an additional roughly linear performance increase orthogonally to the SIMD acceleration. SIMD speed-up is less significant for multi-channel processes --- aligning with the addition of less computationally intense parton configurations --- but multichannel speed-up is now consistently linear with the number of simultaneous executions, consistent with the significant increase in number of channels and phase-space regions to sample. The separation between Fortran and sequential C++ in the many-events limit has been significantly lessened, and the greater-than-2 acceleration when going from one to two simultaneous executions in the bottom plots has completely disappeared. An additional observation which could already be seen in single-channel tests but is especially apparent for multi-channel processes is that Fortran benefits from multiprocessing significantly more than \cudacpp{}, performing on par with SSE4 instructions for simultaneous scheduled executions equal to and greater than the twelve available cores.

\subsection{SIMT GPU results}
\label{sec:evgen_gpu}

\begin{figure}[t]
    \centering
    \includegraphics[width=0.99\linewidth]{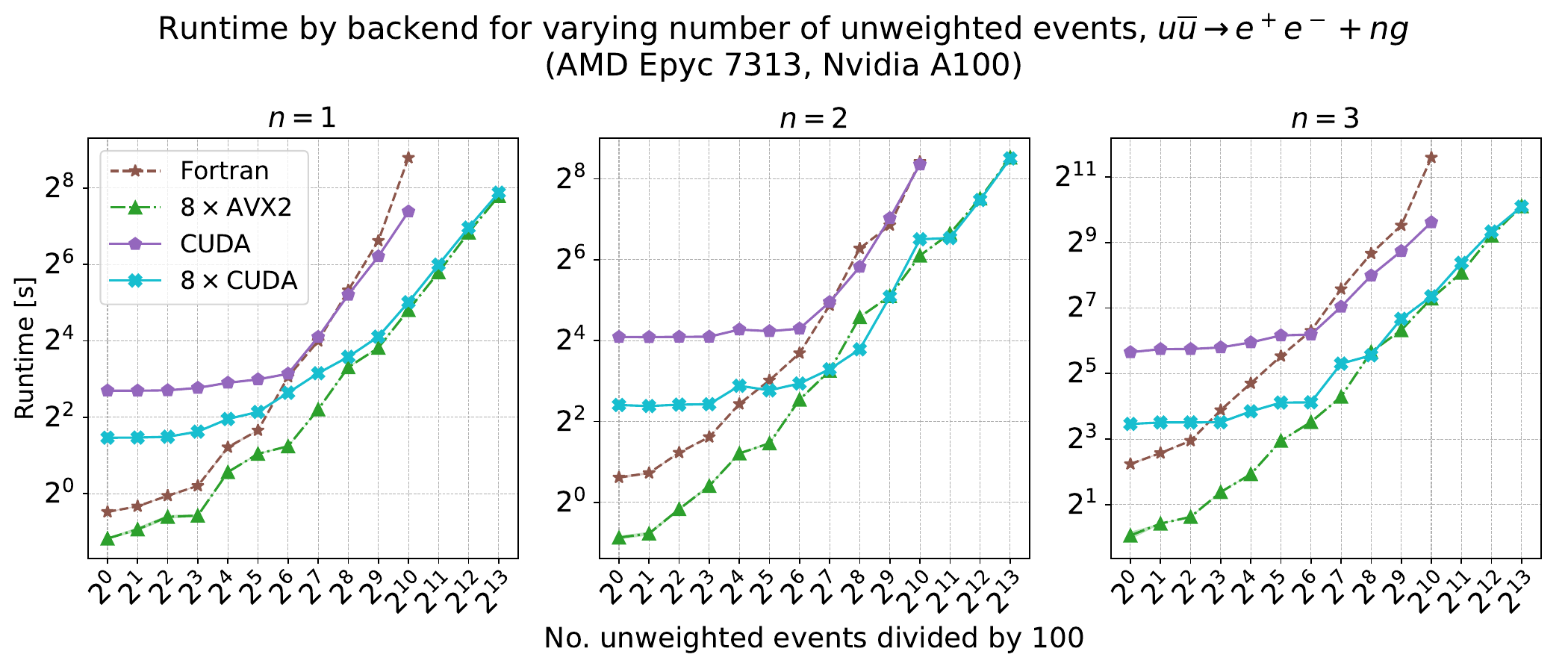}
    \caption{Total runtime to generate unweighted events for the process $u\overline{u}\to e^+ e^- + ng$ for $1 \leq n \leq3$ at LO in \textsc{MadEvent} with fixed number of simultaneous executions and varying number of unweighted events with and without GPU offloading on an Nvidia A100, as well as comparisons to the best available SIMD instructions (AVX2) on the host AMD Epyc 7313 CPU. Fortran and standard CUDA backends are run with a single scheduled execution; the $8\times$ in the labels for AVX2 and the alternative CUDA runs denote that they were run with eight simultaneous executions scheduled by the Python driver. Points are given by the mean runtime of five subsequent runs and standard deviations are highlighted albeit generally too small to be visible.}
    \label{fig:DY_single_gpu}
\end{figure}

\begin{figure}[t]
    \centering
    \includegraphics[width=0.99\linewidth]{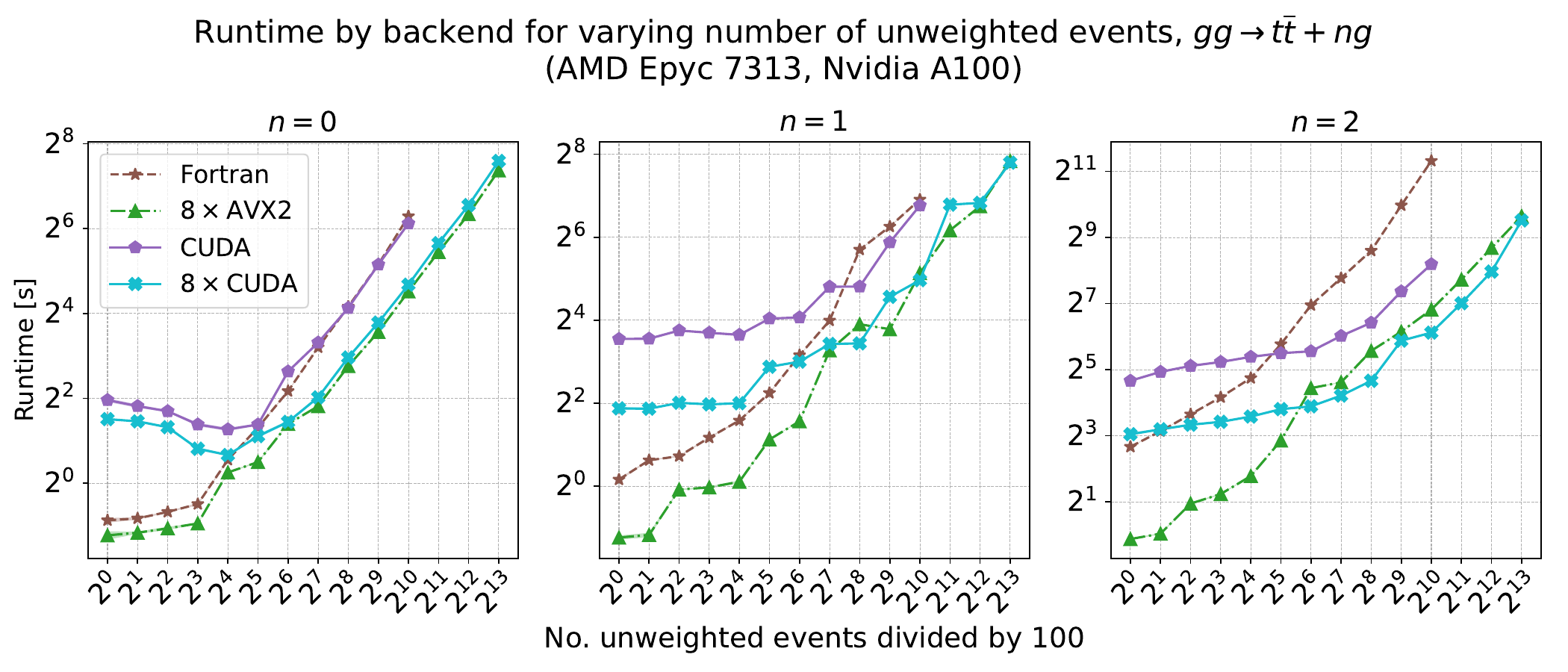}
    \caption{Total runtime to generate unweighted events for the process $gg\to t\overline{t} + ng$ for $0 \leq n \leq2$ at LO in \textsc{MadEvent} with fixed number of simultaneous executions and varying number of unweighted events with and without GPU offloading on an Nvidia A100, as well as comparisons to the best available SIMD instructions (AVX2) on the host AMD Epyc 7313 CPU. Fortran and standard CUDA backends are run with a single scheduled execution; the $8\times$ in the labels for AVX2 and the alternative CUDA runs denote that they were run with eight simultaneous executions scheduled by the Python driver. Points are given by the mean runtime of five subsequent runs and standard deviations are highlighted albeit generally too small to be visible.}
    \label{fig:QCD_single_gpu}
\end{figure}

Finally, it is time to consider GPU event generation. Since GPU execution is generally latency-bound, the number of evaluated events per execution can have a significant impact in a non-uniform way (too few events and latency will throttle the throughput; too many and individual executions will run for exceptionally long times), and since it is implausible for individual channels in a multi-channel processes to occupy the full GPU it is important to optimise the number of simultaneous executions so as to make full use of the GPU without overcommitting the GPU memory (which will generally crash the program). The optimal runtime configuration will vary from process to process; for simplicity, results presented below will be fixed to some standard values.

Just as in \cref{sec:evgen_simd}, the processes considered here are Drell-Yan and $t\overline{t}$-production in the SM at LHC energies, for single-channel and multi-channel variants of these processes. Runtime measurements are taken for generating a varying amount of unweighted events using gridpacks with the maximum number of unweighted events set to the default value of 2 500 and the number of simultaneous executions fixed at 1 and 8. Tests are run on an Nvidia A100 (same as in \cref{sec:standalone_gpu}) with an AMD Epyc 7313 host CPU, and CUDA runtimes are compared to Fortran with single-core execution and AVX2 with 8 simultaneous scheduled \textsc{MadEvent} executions.

\begin{figure}[t]
    \centering
    \includegraphics[width=0.99\linewidth]{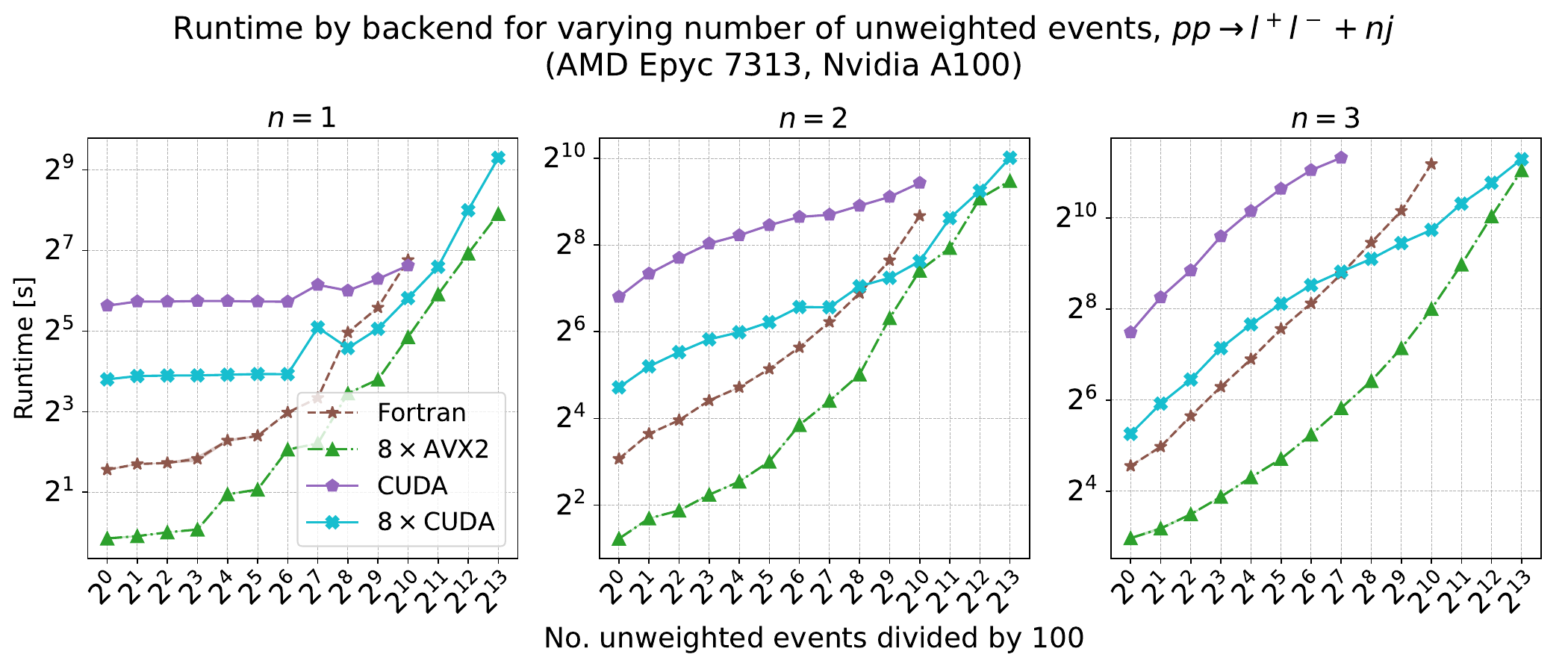}
    \caption{Total runtime to generate unweighted events for the process $pp\to l^+ l^- + nj$ for $1 \leq n \leq3$ with $j$ any massless QCD jet at LO in \textsc{MadEvent} with fixed number of simultaneous executions and varying number of unweighted events with and without GPU offloading on an Nvidia A100, as well as comparisons to the best available SIMD instructions (AVX2) on the host AMD Epyc 7313 CPU. Fortran and standard CUDA backends are run with a single scheduled execution; the $8\times$ in the labels for AVX2 and the alternative CUDA runs denote that they were run with eight simultaneous executions scheduled by the Python driver. Points are given by the mean runtime of five subsequent runs and standard deviations are highlighted albeit generally too small to be visible. Single-process CUDA was cut off after the measurement for 12 800 events for $n=3$ due to excessive runtimes.}
    \label{fig:DY_multi_gpu}
\end{figure}

\begin{figure}[t]
    \centering
    \includegraphics[width=0.99\linewidth]{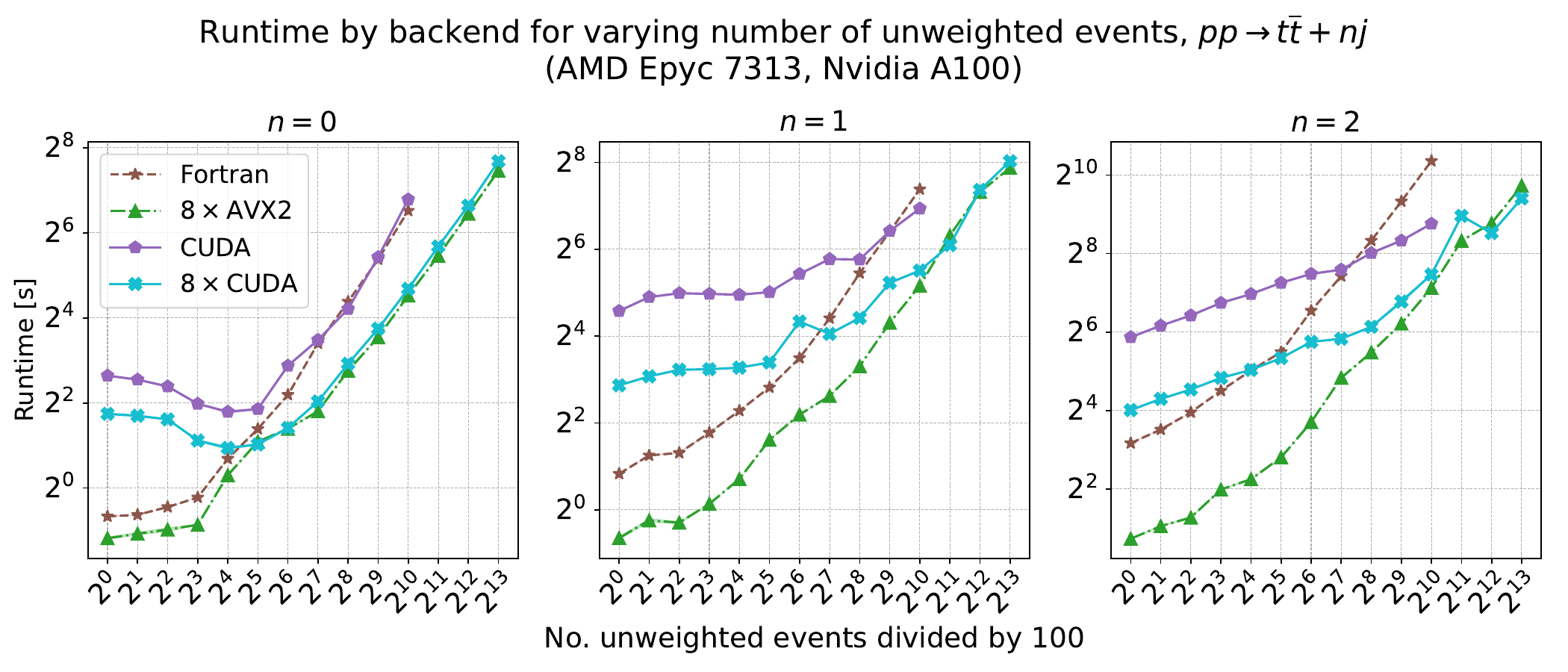}
    \caption{Total runtime to generate unweighted events for the process $pp\to t \overline{t} + nj$ for $0 \leq n \leq2$ with $j$ any massless QCD jet at LO in \textsc{MadEvent} with fixed number of simultaneous executions and varying number of unweighted events with and without GPU offloading on an Nvidia A100, as well as comparisons to the best available SIMD instructions (AVX2) on the host AMD Epyc 7313 CPU. Fortran and standard CUDA backends are run with a single scheduled execution; the $8\times$ in the labels for AVX2 and the alternative CUDA runs denote that they were run with eight simultaneous executions scheduled by the Python driver. Points are given by the mean runtime of five subsequent runs and standard deviations are highlighted albeit generally too small to be visible.}
    \label{fig:QCD_multi_gpu}
\end{figure}

Runtimes as a function of the number of unweighted events for the single-channel processes $u \overline{u} \to e^+ e^- + (n+1)g$ and $g g \to t \overline{t} + ng$ for $0\leq n \leq 2$ are shown in \cref{fig:DY_single_gpu,fig:QCD_single_gpu}, respectively. Fortran and AVX2 shown the exact same behaviour as in \cref{sec:evgen_simd}, while GPU offloading has an approximately constant runtime for smaller numbers of unweighted events but eventually grows into a ``linear regime'' where $t = \mathcal{O}(\text{no. unweighted events})$ and the runtime roughly doubles as the number of requested unweighted events doubles. For the relatively simple single-channel processes shown here, the limiting factor even for $n=2$ is latency, rather than GPU throughput. Consequently, the choice to generate only up to 2 500 unweighted events per execution makes it impossible to fully utilise the Nvidia A100 used here for both single- and multi-process CUDA. Only in the high-complexity many-events limit does single-execution CUDA outperform the Fortran baseline. Similarly, running AVX2 and CUDA with the same number of simultaneous executions have very similar trends in the many-events limit. 

Going to the multi-channel processes $pp \to l^+ l^- + (n+1)j$ and $pp \to t \overline{t} + nj$ ---  with $j$ any massless QCD jet and $0 \leq n \leq 2$ --- shown in \cref{fig:DY_multi_gpu,fig:QCD_multi_gpu}, respectively, the trends for the CUDA lines change. It is particularly apparent in \cref{fig:DY_multi_gpu}, but also in the rightmost plot in \cref{fig:QCD_multi_gpu}, that CUDA runtime is no longer constant in the few-events region. The reason is simple: these multi-channel processes include many more integration channels\footnote{The phase-space grid is computationally split into separate input parameters for the \textsc{MadEvent} executable, meaning each parton configuration will end up with several integration channels.}, and each integration channel will require a minimum of one execution \textit{if} any unweighted events are randomly chosen to be in this region. As the number of unweighted events increases, so does the probability that any given integration channel will be sampled, approaching 100\% in the many-events limit. Thus, long before reaching the previously mentioned ``linear regime'' the GPU runtime will nevertheless increase as the number of necessary executions does. Notice e.g. that single-process CUDA was stopped at $12 800$ events in the rightmost plot of \cref{fig:DY_multi_gpu} due to the excessive runtime despite clearly approaching a plateau prior to the linear regime. Nevertheless, GPU offloading is, at worst, on par with CPU runtimes in the many-events limit.

\begin{figure}[t]
    \centering
    \includegraphics[width=0.45\linewidth]{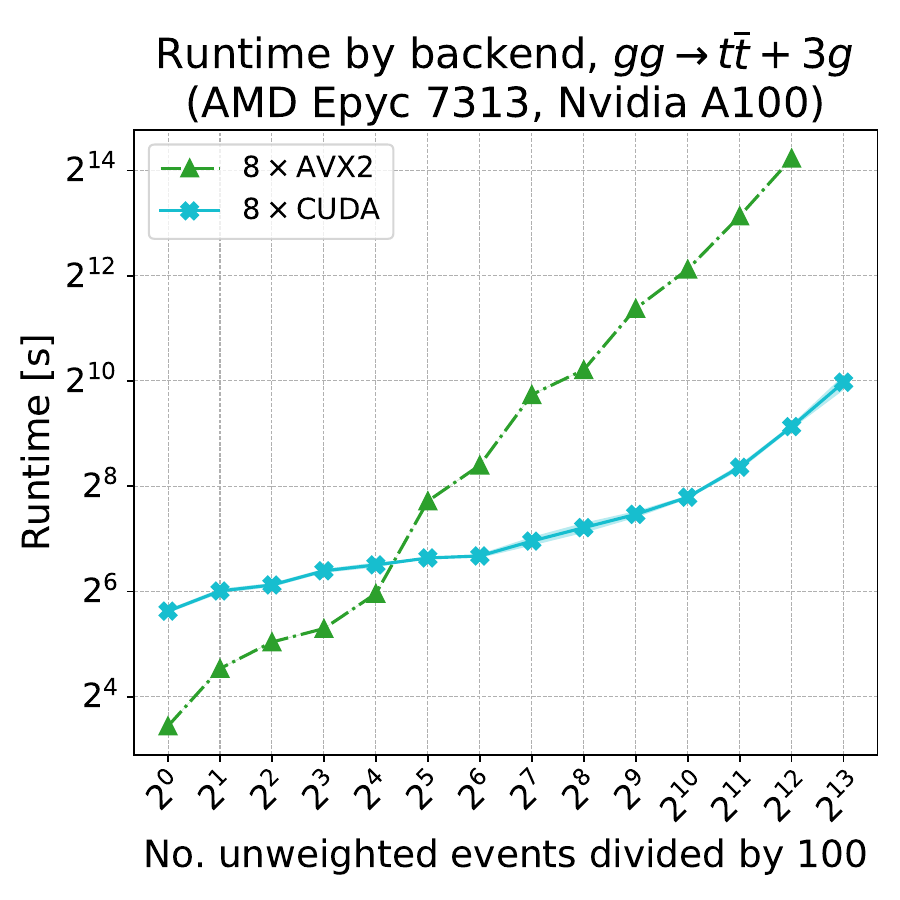}
    \includegraphics[width=0.45\linewidth]{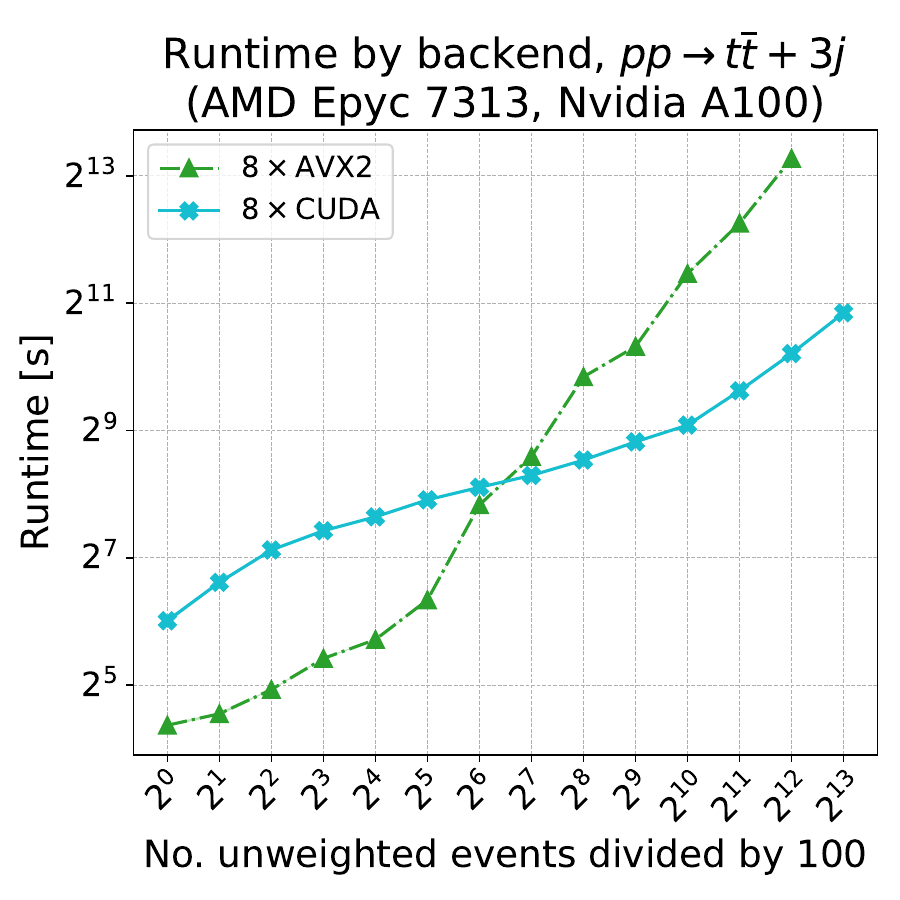}
    \caption{Total runtime to generate unweighted events for the single- and multi-channel processes $gg\to t\overline{t} +3g$ and $p p \to t \overline{t} + 3j$ with $j$ any massless QCD jet at LO in \textsc{MadEvent} with eight simultaneous executions and varying number of unweighted events with the best available SIMD instructions (AVX2) on the AMD Epyc 7313 host CPU and with GPU offloading on an Nvidia A100. Tests were run with eight simultaneous executions scheduled by the Python driver (gridpack runtime flag \texttt{-p 8}). Points are given by the mean runtime of five subsequent runs and standard deviations are highlighted albeit generally too small to be visible. The maximal number of unweighted events generated by any one execution has been limited to 1 000 (gridpack runtime flag \texttt{-m 1000}).}
    \label{fig:QCD_3j_gpu}
\end{figure}

Of course, GPU acquisition cannot be justified with event generation runtimes on par with a server-grade CPU. While the GPU runtimes above could be optimised somewhat with runtime parameters, the fundamental restriction lies in the relative simplicity in these calculations --- the runtime on the GPU itself is negligible, but the latency in off-host communication severely limits the possible speed-up. For single-channel processes this can be overcome by maximising the number of events generated on the device per call and limiting the number of communication instances. This solution does not apply to multi-channel processes, where the number of integration channels is the more significant factor in determining the number of necessary executions (as can be seen in the single-process CUDA line on the rightmost plot of \cref{fig:DY_multi_gpu}). Furthermore, most of these channels will end up being computationally cheap, leaving only a few scheduled executions to handle the most computationally challenging channels --- in the end, this would end up underutilising the GPU instead. 

To truly see the benefit of GPU offloading more complex processes must be considered --- the very processes that are bottlenecks. In \cref{fig:QCD_3j_gpu}, event generation runtimes for single- and multi-channel $t\overline{t}$-production with three additional final state jets are shown for an increasing amount of unweighted events. Since these processes require extreme runtimes without hardware acceleration, here comparisons are taken only between multi-process AVX2 and CUDA.

Roughly the same trends can be seen across both plots in \cref{fig:QCD_3j_gpu}; in the many-events limit, AVX2 reaches an amplitude-dominated linear regime --- runtime roughly doubling as the requested number of unweighted events does --- while GPU offloading still remains just short of linear runtime growth, especially in the right-hand plot multi-channel process. In the few-events region, CUDA runtimes for the single-channel process $gg \to t \overline{t} + 3g$ remains roughly constant while for the multi-channel process $pp \to t \overline{t} + 3j$ it increases significantly with increasing events. This reinforces the analysis of GPU off-loading bottlenecks: each individual channel is largely latency-bound, but for processes with many channels not all of them are sampled when generating only few unweighted events. Eventually most to all channels are sampled, at which point the runtime growth will stagnate before eventually increasing again as single executions no longer provide sufficiently many unweighted events per channel.

While Fortran tests are omitted from \cref{fig:QCD_3j_gpu} due to expected exceptional runtimes, a lower bound of speed-up in the AVX2 case can be estimated from the results for $n=2$ in \cref{fig:QCD_single_channel,fig:QCD_multi_channel}; for $g g \to t \overline{t} +2g$, AVX2 with \texttt{-p 8} is roughly $16\times$ faster at generating 64 000 events than single-process Fortran, whereas for $p p \to t \overline{t} +2j$ it is only about $8\times$ faster. Taking these as lower bounds for the runtime ratios and noting that \cref{fig:QCD_3j_gpu} indicates that GPU offloading provides an additional speed-up of $32\times$ ($8\times$) for $gg \to t \overline{t} +3g$ ($p p \to t \overline{t} +3j$) in the many-events limit, GPU offloading is at a minimum  $512\times$ ($64\times$) faster than single-process Fortran.

\begin{figure}[t]
    \centering
    \includegraphics[width=0.95\linewidth]{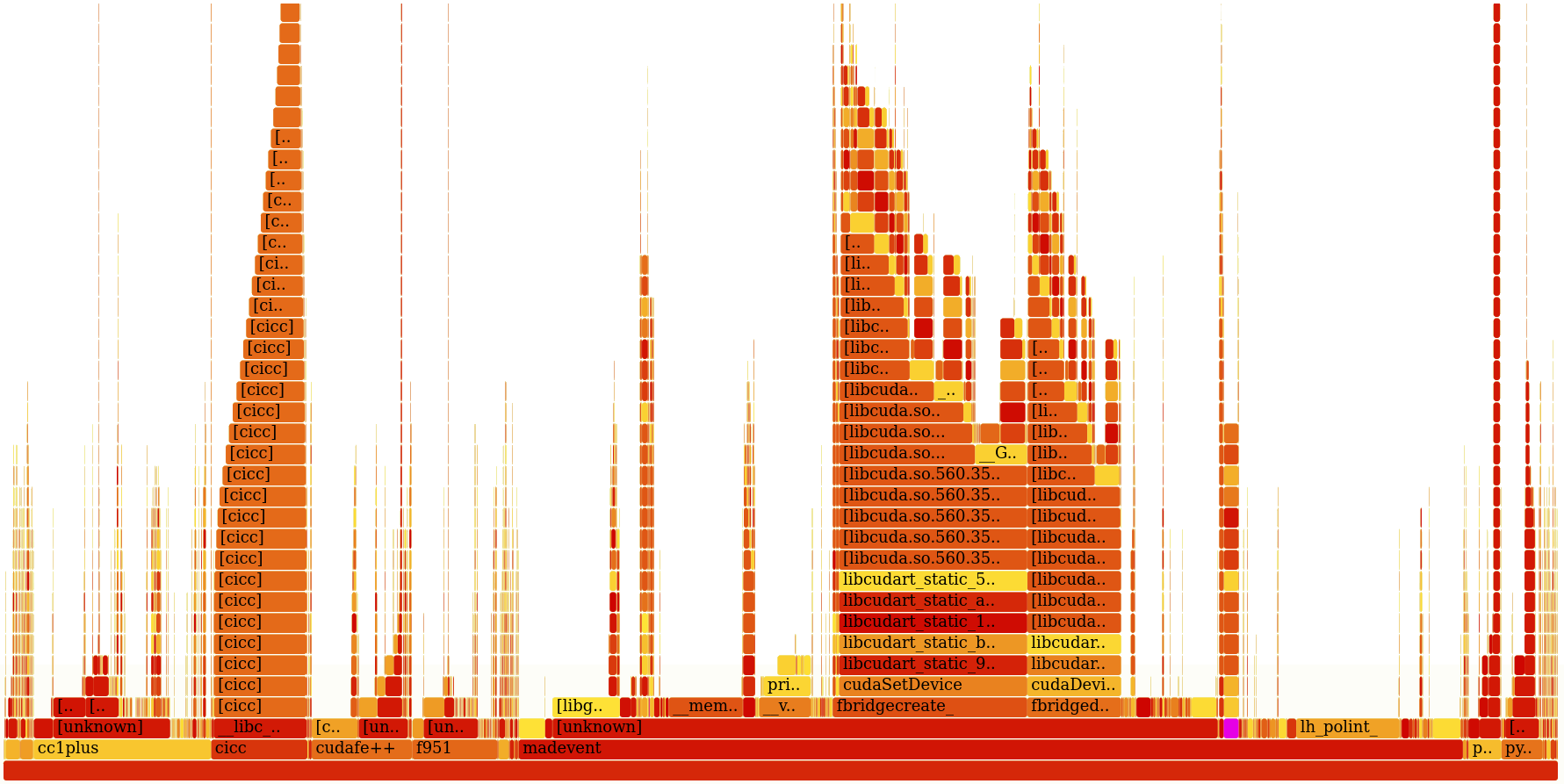}
    \caption{FlameGraph \cite{brendangreggFlameGraphs} of \cudacpp{} event generation with GPU offloading on an Nvidia A100 for the process $gg\to t \overline{t} \, gg$, generating $10\, 000$ unweighted events. Compare to \cref{fig:flamegraph}, which provides the equivalent profile for Fortran event generation. The mixture of executables and libraries written in different languages makes stack traces difficult to provide, but everything on top of the \texttt{madevent} Python driver is part of event generation. The function \texttt{computeMatrixElements}, equivalent to \texttt{smatrix} in Fortran, has been highlighted but makes up only  $1.56\%$ ($0.95\%$) of \texttt{madevent} (total) runtime and can barely be identified even when highlighted.}
    \label{fig:flamegraph_gpu}
\end{figure}

\Cref{fig:DY_single_gpu,fig:QCD_single_gpu,fig:DY_multi_gpu,fig:QCD_multi_gpu,fig:QCD_3j_gpu} all together serve to illustrate both the limitations and benefits of GPU offloading: for computationally simple processes there is little to no benefit in GPU offloading when compared to better utilisation of available CPU resources; for sufficiently complicated processes, though, GPU offloading can provide speed-up over a well-utilised CPU by more than an order of magnitude. Notice that in \cref{fig:QCD_3j_gpu} AVX2 runtimes are lower for the multi-channel process than for the single-channel process, as the latter is the computationally most expensive part of the former, whereas CUDA runtimes increase with the additional necessary executions required by the multi-channel process. A scheme where GPU off-loading is only scheduled for the computationally heaviest subprocesses could plausibly provide the best aspects of both host-side hardware acceleration and GPU offloading.

A final illustration of the speed-up provided by GPU offloading is given in \cref{fig:flamegraph_gpu} which shows an equivalent FlameGraph to \cref{fig:flamegraph} for \cudacpp{} event generation for the same process and the same number of generated unweighted events. The mixture of programming languages and how they interface limits the availability of stack traces, but several things are immediately obvious: compilation is now a far larger part of total runtime, as seen by \texttt{f951} alongside several more compiler symbols now being clearly legible on the left side of the FlameGraph; and a significant \texttt{madevent} runtime fraction is now spent in the Fortran-side bridge management routines \texttt{fBridgeCreate} and \texttt{fBridgeDelete} described in \cref{sec:technicalevgen}. No samples were taken in \texttt{fBridgeSequence}, the Fortran-side subroutine that actually launches CUDA kernels on the GPU, but many samples were taken in the resulting CUDA function call \texttt{computeMatrixElements} --- the \cudacpp{} equivalent of the \texttt{smatrix} routine in Fortran. In \cref{fig:flamegraph} \texttt{smatrix1} makes up $92.7\%$ of \texttt{madevent} runtime, whereas in \cref{fig:flamegraph_gpu} \texttt{computeMatrixElements} makes up only $1.56\%$. Clearly, the actual GPU helicity amplitudes are several orders of magnitude faster than the corresponding Fortran ones running on a CPU; the limitation comes in host-device latency, as is shown by the \texttt{fBridge} routines combined taking $35.5\%$ of \texttt{madevent} runtime.

\section{Conclusion and outlook}
\label{sec:conclusions}

The \cudacpp{} plugin for \madgraph{}, which replaces the ALOHA-generated helicity amplitude routines in Fortran with equivalent ones in templated C++ and CUDA with explicit event-level data parallelism implemented, speeds up helicity amplitude evaluations by the theoretical linear factor expected based on the SIMD register size when using SIMD instructions and the number of available threads when multithreading, and even more with SIMT GPU offloading on HPC-grade GPUs. However, when interfacing \cudacpp{}-generated helicity amplitudes with \textsc{MadEvent} event generation executables the speed-up is limited by remaining sequential parts in the executable; nevertheless, the speed-up using SIMD instructions in general behaves according to Amdahl's law based on process complexity while for sufficiently complex processes GPU offloading can make helicity amplitudes negligible when compared with the original Fortran implementation. Essentially, when using \cudacpp{} helicity amplitudes are no longer the primary bottleneck in LO event generation. \cudacpp{} is available publicly on GitHub \cite{Valassi_Roiser_Hageboeck_Smith_2020} but can more easily be installed through the \mg{} command line interface with the command ``\texttt{install cudacpp}'' (see \cref{appendix:manual}).

Due to the perfect scaling of \cudacpp{} amplitudes, near-future optimisation developments consider practical restrictions in \cudacpp{} usage rather than amplitude code itself; output code is highly centralised, making compilation times far longer in \cudacpp{} than corresponding default Fortran code. This compilation slowdown is further exaggerated by \cudacpp{} parton grouping --- or rather, the lack thereof. To ensure exact lockstep between iterations within the same executable, processes that would be compiled into separate branches of the same executable with default \mg{} code are separated into distinct compiled executables. This also increases the number of distinct integration channels to sample in \textsc{MadEvent} and consequently the number of scheduled jobs needed for event generation. Work on this is ongoing both with respect to simplifying the compilation of individual executables (by simplifying the output code structure to minimise the number of simultaneously optimised code sections) and with respect to the number of distinct executables (by developing a new parton grouping scheme that minimises branching within a parton group).

Accelerating LO event generation further requires speeding up remaining sequential parts of the code. In \mg{}, there remains no singular obvious bottleneck past the scattering amplitudes, but other groups are working on e.g. porting parton distribution functions to GPUs \cite{Carrazza:2020qwu,gitlabFilesKokkos_version} or using neural importance sampling to optimise phase-space sampling on both CPUs and GPUs \cite{Bothmann:2020ywa,Heimel:2022wyj,Heimel:2023ngj,Heimel:2024wph,bothmann2025efficientmanyjeteventgeneration}. There is an ongoing effort in integrating \cudacpp{} and such developments with the hope that GPU offloading can become more practical outside of the very high complexity limit by performing larger sections of the program on the GPU and especially \textit{staying} on the GPU for a larger runtime fraction, mitigating the latency-driven overhead.

Going beyond LO, there are several different and distinct projects in applying hardware acceleration to NLO event generation. In the context of \mg{} and \cudacpp{}, there are concurrent efforts in developing a multi-event interface for the NLO codebase (which in \mg{} is largely distinct from LO) and in extending \cudacpp{} code generation to account for the differences in amplitude code generated for NLO tree-level amplitudes \cite{Wettersten:2023ekm,Wettersten:2025hrb}. Simultaneously, other groups are working on porting loop libraries to data-parallel hardware architectures. These developments are largely independent of one another, with the optimistic expectation that combining them into a single event generation package should be relatively simple.

In summary, \cudacpp{} provides significant speed-up for LO event generation in \mg{} --- using SIMD CPUs, maximal theoretical acceleration is achieved, while using SIMT GPUs sufficiently complicated processes can be sped up by an order of magnitude compared even to optimal CPU usage with SIMD instructions. The immediate short-term priority in terms of \cudacpp{} development concerns the simplification of the generated code structure to circumvent issues to do with compilation times and number of distinct integration channels, as well as integrating \cudacpp{} and other projects to accelerate further aspects of event generation. In the longer term, the aim is to port also NLO event generation to data-parallel architectures, enabling hardware-accelerated event generation even at higher order precisions.

\section*{Acknowledgements}
We thank all the contributors to the \cudacpp{} project, as well as contributors to all adjacent projects whose work has directly and indirectly impacted \cudacpp{} development\footnote{For a full list of \cudacpp{} project members, see \href{https://github.com/orgs/madgraph5/people}{https://github.com/orgs/madgraph5/people}.}. We also thank all \mg{} authors, past and present. Computational resources were provided by the Calcul Intensif et Stockage de Masse (CISM) technological platform. The \cudacpp{} project has in part been supported by CERN openlab. DM, SR, and ZW have also in part been supported by the Next Generation Triggers project jointly hosted by CERN and the ATLAS and CMS experiments, which is funded by the Eric and Wendy Schmidt Fund for Strategic Innovation.

\clearpage

\appendix
\section{\cudacpp{} installation and usage manual}
\label{appendix:manual}

This appendix is intended to serve as a guide on how to install and use the \cudacpp{} plugin. Technical details will be kept light here --- more extensive and up-to-date information can be found on the \cudacpp{} GitHub wiki \cite{Valassi_Roiser_Hageboeck_Smith_2020}, and if something is not detailed there or in previous \cudacpp{} conference proceedings \cite{Valassi:2021ljk,Valassi:2022dkc,Valassi:2023yud,Hageboeck:2023blb,Valassi:2025xfn} the \cudacpp{} development team can be contacted directly.

To install and run \cudacpp{}, consult \cref{app:install,app:codegen} respectively. More extensive options for \cudacpp{} event generation are detailed in \cref{app:runcard}. It is assumed that the reader is already familiar with \madgraph{} and its command line interface (CLI). If you are not, the \mg{} Wiki \cite{uclMadGraph} or Launchpad site \cite{launchpadMadGraph5_aMCNLOLaunchpad} serve as a good introduction.

\subsection{Installation}
\label{app:install}

Installing \cudacpp{} from within \mg{} can be done with a single command in the CLI:
\begin{lstlisting}[language=terminal]
install cudacpp
\end{lstlisting}
\noindent which will automatically download the latest validated \cudacpp{} release and unpack it in the \texttt{/PLUGIN/} subdirectory, allowing immediate access to \cudacpp{} output routines. To install a particular \cudacpp{} release, the flag \texttt{\color[HTML]{1027ad}-{}-cudacpp\_tarball=} can be appended followed by the URL of a \cudacpp{} tarball, as listed under ``Releases'' on the \cudacpp{} GitHub page.

Should you wish to install the development version of \cudacpp{}, it can be cloned with \texttt{\color[HTML]{228B22}git}:
\begin{lstlisting}[language=terminal]
git clone https://github.com/madgraph5/madgraph4gpu.git
\end{lstlisting}
\noindent after which the a symbolic link to the \cudacpp{} interface directory needs to be created within the \texttt{/PLUGIN/} subdirectory of the desired \mg{} release as described below. The latest validated \mg{} version is provided as a submodule of \cudacpp{} and can be cloned alongside the main repository with the git flag \texttt{\color[HTML]{1027ad}{-}{-}recurse-submodules}.

Assuming \mg{} was cloned as a submodule, \cudacpp{} can be installed with
\begin{lstlisting}[language=terminal]
cd madgraph4gpu/MG5aMC/mg5amcnlo/PLUGIN/
ln -s ../../MG5aMC_PLUGIN/CUDACPP_OUTPUT/ CUDACPP_OUTPUT
\end{lstlisting}
which creates a symbolic link to the directory \texttt{../../MG5aMC\_PLUGIN/CUDACPP\_OUTPUT/} which, in turn, is a symbolic link to the directory \texttt{CUDACPP\_SA\_OUTPUT} hidden away in the \texttt{CODEGEN} subdirectory of \texttt{epochX/cudacpp}, sitting in the repository's main directory. Installing \cudacpp{} in a different installation of \mg{} can be done similarly by switching the path in the symbolic link above with the corresponding one to reach this directory from that \mg{} installation's plugin subdirectory.

\subsection{Code generation}
\label{app:codegen}

\subsubsection{Standalone}

\cudacpp{} provides three alternative \texttt{output} methods to \mg{}. The first one is the standalone output described in \cref{sec:technicalstandalone}, which can be generated using the command
\begin{lstlisting}[language=terminal]
generate PROC
output standalone_cudacpp
\end{lstlisting}
although unlike the Fortran standalone output \texttt{\color[HTML]{1027ad}standalone\_cudacpp} cannot be launched directly from the \mg{} CLI as the executable takes three mandatory runtime arguments in the number of blocks per kernel call, the number of threads per block, and the number of iterations to run. The former two are only individually relevant when compiling for SIMT GPUs; on SIMD CPUs, only the product of them has runtime significance.

To actually run the standalone application, \texttt{\color[HTML]{228B22}cd} into the \texttt{PROC/SubProcesses} directory and from there into (one of) the specific parton configuration(s), each of which is given as a subdirectory of the form \texttt{P1\_Sigma\_...} with the subdirectories named for each distinct parton configuration. Note that no parton grouping is done at this level, meaning subprocesses with identical helicity amplitude routines will nevertheless be generated as distinct codes. Once in one of these directories, run e.g.
\begin{lstlisting}[language=terminal]
make (BACKEND=cppauto FPTYPE=m USEOPENMP=0)
./check_cpp.exe 32 32 32
\end{lstlisting}
where the default compilation options have been included in parenthesis and do not need to be written explicitly, but can be chosen differently. Note that the executable will be named \texttt{\color[HTML]{228B22}check\_cuda.exe} when compiling for CUDA and \texttt{\color[HTML]{228B22}check\_hip.exe} for HIP. The second line launches the \cudacpp{} standalone application and runs 32 iterations of $32\times32 = 1\,024$ events each --- for more extensive options, run \texttt{{\color[HTML]{228B22}./check\_cpp.exe} {\color[HTML]{1027ad}-h}}.

\subsubsection{Event generation}

To use \cudacpp{} event generation, the relevant outputs are \texttt{madevent\_simd} for SIMD CPUs and \texttt{madevent\_gpu} for SIMT GPUs, which can be used as
\begin{lstlisting}[language=terminal]
generate PROC
output madevent_simd
launch
\end{lstlisting}
or equivalently for \texttt{\color[HTML]{1027ad}madevent\_gpu}, where the event generation execution \textit{can} be launched directly from the \mg{} CLI unlike the standalone output. These outputs will generate and output identical code since the hardware backend is defined only at compile time, although the surrounding input parameters (as described in \cref{app:runcard}) will be set to chosen defaults for the corresponding expected hardware. This will automatically compile and run event generation for the requested process on SIMD CPUs or Nvidia GPUs.

\subsection{\texttt{run\_card.dat} arguments}
\label{app:runcard}

Several options in the \texttt{run\_card.dat} file can be used to customise the \cudacpp{} compilation. There are two additional flags that can be defined besides the ones already provided by \mg{}:
\begin{itemize}
    \item \texttt{\color[HTML]{1027ad}floating\_type}: The desired floating point type. Supported options are \texttt{\color[HTML]{1027ad}d} (FP64), \texttt{\color[HTML]{1027ad}f} (FP32), and \texttt{\color[HTML]{1027ad}m} (mixed). Default is \texttt{\color[HTML]{1027ad}m}. For further details, consult \cref{sec:precision}.
    \item \texttt{\color[HTML]{1027ad}cudacpp\_backend}: Which hardware architecture to compile the executables for. 
    \begin{itemize}
        \item SIMD CPUs: Supported options are \texttt{\color[HTML]{1027ad}cpp}/\texttt{\color[HTML]{1027ad}cppauto} (which will automatically detect the best available SIMD instruction set), \texttt{\color[HTML]{1027ad}cppsse4} (SSE4), \texttt{\color[HTML]{1027ad}cppavx2} (AVX2), \texttt{\color[HTML]{1027ad}cpp512y} (AVX-512 compiled for 256-bit registers), and \texttt{\color[HTML]{1027ad}cpp512z} (AVX-512 compiled for 512-bit registers). Default is \texttt{\color[HTML]{1027ad}cpp}/\texttt{\color[HTML]{1027ad}cppauto}.
        \begin{itemize}
            \item Note that \texttt{\color[HTML]{1027ad}cppauto} will default to \texttt{\color[HTML]{1027ad}cpp512y} if AVX-512 instructions are detected. If you know your machine has two AVX-512 units per core, instead explicitly use \texttt{\color[HTML]{1027ad}cpp512z}.
        \end{itemize}
        \item SIMT GPUs: Supported options for SIMT GPUs are \texttt{\color[HTML]{1027ad}cuda} (CUDA, Nvidia GPUs) and \texttt{\color[HTML]{1027ad}hip} (HIP, AMD GPUs). Default is \texttt{\color[HTML]{1027ad}cuda}.
    \end{itemize} 
\end{itemize}

Additionally, several \texttt{run\_card.dat} options available in the upstream \mg{} project are relevant for \cudacpp{} event generation. First are those relating to vectorisation:
\begin{itemize}
     \item \texttt{\color[HTML]{1027ad}vector\_size}: Number $n$ of consecutive events to evaluate in lockstep per parton configuration, i.e. the block size. Should be kept as small as plausible to minimise bias: For SIMD CPUs, the number of FP32 numbers that fit in one SIMD register (i.e. $4-16$ for $128-512$ bit registers) is recommended; for SIMT GPUs, the GPU warp size (i.e. \texttt{\color[HTML]{1027ad}32} for Nvidia GPUs, \texttt{\color[HTML]{1027ad}64} for AMD GPUs) is recommended. Default is \texttt{\color[HTML]{1027ad}16} and \texttt{\color[HTML]{1027ad}32} for SIMD and GPU output modes, respectively.
     \item \texttt{\color[HTML]{1027ad}nb\_warp}: Number $p$ sets of \texttt{\color[HTML]{1027ad}vector\_size} $n$ phase-space points to generate before launching the helicity amplitude kernel. The total number of events scheduled at once (for SIMT GPUs, the grid size) is equal to $p\times n$. For SIMD CPUs, this parameter is superfluous and should be kept to \texttt{\color[HTML]{1027ad}1}. For SIMT GPUs, it should be taken sufficiently large to make proper use of the GPU. A decent choice is \texttt{\color[HTML]{1027ad}512}, scheduling 16 384 events on the GPU for $n=32$. Default is \texttt{\color[HTML]{1027ad}1} and \texttt{\color[HTML]{1027ad}512} for SIMD and GPU output modes, respectively.
\end{itemize}
Then are the variables related to phase-space handling, for which \cudacpp{} has restrictions:
\begin{itemize}
    \item \texttt{\color[HTML]{1027ad}nhel}: Defines whether helicities should be summed (\texttt{\color[HTML]{1027ad}0}) or sampled (\texttt{\color[HTML]{1027ad}1}) during event generation. As mention in the main text, \cudacpp{} does not support helicity sampling at present making the default value of \texttt{\color[HTML]{1027ad}nhel=0} necessary.
    \item \texttt{\color[HTML]{1027ad}sde\_strategy}: Which integration strategy to use; standard single diagram enhancement (\texttt{\color[HTML]{1027ad}1}) \cite{Maltoni:2002qb} or propagator-specific enhancement (\texttt{\color[HTML]{1027ad}2}) \cite{Mattelaer:2021xdr}. Only the classic diagram enhancement method \texttt{\color[HTML]{1027ad}sde\_strategy=1} is supported by \cudacpp{} and will as such always be selected by default.
\end{itemize}

\clearpage

\bibliography{references}
\end{document}